\documentclass[aps,twocolumn,showpacs]{revtex4}
\def\[{\left\lbrack}
\def\]{\right\rbrack}

\def\({\left(}
\def\){\right)}
\newcommand{\be}{\begin{equation}}
\newcommand{\ee}{\end{equation}}
\newcommand{\ea}{\end{eqnarray}}
\newcommand{\ba}{\begin{eqnarray}}

\usepackage{graphicx}
\usepackage{colordvi}

\begin{document}

\title{Can noncommutativity affect the whole history of the Universe?}

\author{G. A. Monerat\footnote{E-mail: monerat@uerj.br}}

\affiliation{Departamento de Modelagem Computacional,\\
Instituto Polit\'{e}cnico do Rio de Janeiro,\\
Universidade do Estado do Rio de Janeiro,\\
Rua Bonfim, 25 - Vila Am\'{e}lia - Cep 28.625-570,
Nova Friburgo, RJ, Brazil.}

\author{E. V. Corr\^{e}a Silva\footnote{E-mail: evasquez@uerj.br}}

\author{C. Neves\footnote{E-mail: cliffordneves@uerj.br}}

\affiliation{Departamento de Matem\'{a}tica, F\'\i sica e Computa\c{c}\~{a}o, \\
Faculdade de Tecnologia, \\ 
Universidade do Estado do Rio de Janeiro,\\
Rodovia Presidente Dutra, Km 298, P\'{o}lo
Industrial,\\
CEP 27537-000, Resende, RJ, Brazil.}

\author{G. Oliveira-Neto\footnote{E-mail: gilneto@fisica.ufjf.br}}

\author{L. G. Rezende Rodrigues\footnote{E-mail: luiguilherme 7@hotmail.com}}

\author{M. Silva de Oliveira\footnote{E-mail: monalisa-silva@hotmail.com}}

\affiliation{Departamento de F\'{\i}sica, \\
Instituto de Ci\^{e}ncias Exatas, \\
Universidade Federal de Juiz de Fora,\\
CEP 36036-330 - Juiz de Fora, MG, Brazil.}

\begin{abstract}
We study a classical, noncommutative (NC), Friedmann-Robertson-Walker cosmological model.
The spatial sections may have positive, negative or zero constant curvatures. 
The matter content is a generic perfect fluid. The initial noncommutativity between some canonical 
variables is rewritten, such that, we end up with commutative variables and a NC parameter.
Initially, we derive the scale factor dynamic equations for the general situation, without 
specifying the perfect fluid or the curvature of the spatial sections. Next, we consider two
concrete situations: a radiation perfect fluid and dust. We study all possible scale factor 
behaviors, for both cases. We compare them with the corresponding commutative cases and one with the other. 
We obtain, some cases, where the NC model predicts a scale factor expansion which may describe the present
expansion of our Universe. Those cases are not present in the corresponding commutative models.
Finally, we compare our model with another NC model, where the noncommutativity is between different
canonical variables. We show that, in general, it leads to a scale factor behavior that is different 
from our model.
\end{abstract}

\pacs{04.20.Fy,04.40.Nr,11.10.Nx,98.80.Jk}

%\keywords{cosmology, noncommutativity, universe, accelerated expansion}

\maketitle 

\section{Introduction}
\label{sec:intro}

Almost twenty years ago it was possible to conclude from observational data that
the Universe is going through a phase of accelerated
expansion \cite{expansion}. After that, many hypothesis appeared in the
literature trying to explain this fascinating feature. The different explanations
are divided in two major groups: the ones using general relativity (GR) as the correct
theory to explain the gravitational phenomena and the others that do not use it.
Among the ones in the first group we may mention the models that consider the
Universe uniformly filled with a phantom fluid (the dark energy),
which would drive the accelerated expansion of the Universe under an
equation of state $p/\rho = \alpha < -1/3$, where $\alpha$ is a constant which defines 
the fluid, $p$ is the fluid pressure and $\rho$ its
density \cite{phantom}. In this context, the cosmological
constant is an important candidate to represent the dark energy,
since it acts as a fluid with $p/\rho = -1$ ($\Lambda$CDM model), which is
consistent with observational results \cite{data1}. However, the theoretical value
for the cosmological constant (estimated by the physics of high energy
particles) conflicts with observational data by 30 orders of magnitude
in the energy scale \cite{data2}. Despite of this, the role played by the
cosmological constant at the FRW cosmological model has been
investigated at the quantum level \cite{nivaldo} and interesting results
have been obtained. The positive cosmological constant can take account
of the initial inflationary period and also of a late accelerated
expansion phase, with an in-between decelerated expansion phase. Also,
the past and future cosmological singularities are removed. 
Among the explanations to the present accelerated expansion using
theories other than GR we may mention the ones which
modify the 4-dimensional Einstein-Hilbert action by
adding higher order curvature invariants. In the simplest
examples in this class, the GR action is modified by the addition of
Ricci scalar ($R$) functions. They are the so called f(R) theories \cite{f(R)}.
For a review on different theoretical explanations for the
present accelerated expansion see Ref. \cite{trodden}.

Noncommutative ideas were first introduced, a long time ago, by Snyder \cite{snyder,snyder1}. 
There, the noncommutativity was imposed between the spacetime coordinates 
and his main motivation was to eliminate the divergences in quantum field theory.
Recently, the interest in those ideas of noncommutativity between spacetime coordinates were 
renewed due to some important results obtained in superstring, membrane and $M$-theories 
\cite{banks,connes,chu,schomerus,witten}.
In the past few years, the role played by noncommutativity in cosmological models has been extensively
investigated \cite{NCC}. In this scenario it
was possible to prove that the past and future cosmological
singularities are removed due to noncommutativity \cite{barbosa1} and that
the NC effects are relevant to the entire history of the Universe, for
intermediary times \cite{nelsonP}. The main motivation in order to consider noncommutativity 
in classical cosmology models is the possibility that some residual noncommutative contribution 
may have survived in later stages of our Universe. Based on these ideas some 
researchers have proposed some NC models in classical cosmology in order to explain 
some intriguing results observed by WMAP. Such as a running spectral index of the 
scalar fluctuations and an anomalously low quadrupole of CMB angular power spectrum 
\cite{huang,kim,liu,huang1,kim1}. Another relevant application of NC ideas in 
semi-classical and classical cosmology is the attempt to explain the present accelerated 
expansion of our Universe \cite{pedram,obregon,gil}.

In the present work, we would like to contribute to the investigation on the importance of 
noncommutativity as a possible mechanism to explain the present expansion of the Universe. 
In this way, we study the noncommutative version of a classical
cosmology model. The model has a Friedmann-Robertson-Walker (FRW) geometry, the 
matter content is, initially, a generic perfect fluid and the spatial sections may have 
negative, positive or zero constant curvatures. We work in the Schutz's variational 
formalism \cite{schutz,germano1}. The noncommutativity is
obtained by imposing deformed Poisson brackets between certain canonical variables. 
Initially, we derive the scale factor dynamic equations for the general situation, without 
specifying the perfect fluid or the curvature of the spatial sections. Next, we consider two
concrete situations: a radiation perfect fluid and dust. We study all possible scale factor 
behaviors, for both cases. We compare them with the corresponding commutative cases and one 
with the other. We obtain, some cases, where the NC model predicts a scale factor expansion 
which may describe the present expansion of our Universe. Those cases are not present in the 
corresponding commutative models. The noncommutativity that
we are about to propose is not the typical noncommutativity between usual spatial
coordinates. We are describing FRW models using the Hamiltonian formalism, therefore
their phase spaces are given by the canonical variables and conjugated
momenta: $\{ a, P_a, T, P_T \}$. Then, the noncommutativity we 
are about to propose will be between these phase space variables. Since these variables 
are functions of the time coordinate $t$, this procedure is a generalization of the typical 
noncommutativity between usual spatial coordinates. The noncommutativity between those
types of phase space variables have already been proposed in the literature. At the quantum
level in Refs. \cite{garcia,nelsonP,barbosa1,obregon} and at the classical level in
Refs. \cite{pedram,gil}. Another motivation of the present work is investigating if different 
ways of introducing noncommutativity, in the same cosmological model, modify the predictions 
of the resulting NC models. In this way we present a detailed comparison with another NC model 
\cite{gil}.

In Section \ref{sec:general}, we introduce the noncommutative model for a generic
perfect fluid and derive the coupled system of differential equations for the variables. In
Section \ref{sec:radiation}, we apply the general formalism for the case of a radiation perfect
fluid. We solve the system of differential equations and obtain the scale factor as a function
of the time coordinate and few parameters, including the NC parameter $\gamma$. We analyze all
possible behavior of the solutions, including a comparison with the solutions for the corresponding 
commutative model, paying special attention for those representing expansion. In Section \ref{sec:dust}, 
we repeat the investigation performed in the previous section, this time for the case of a dust perfect 
fluid. We, also, compare the scale factor behavior for the case of a dust perfect fluid with the behavior 
for the case of a radiation perfect fluid.
In Section \ref{sec:comparison}, we present a detailed comparison with another NC model 
\cite{gil}, where the noncommutativity is between different canonical variables. We show that, in general, 
it leads to a scale factor behavior that is different from our model.
Finally, in Section \ref{sec:conclusions}, we comment on the most important results of the present paper.

\section{The noncommutative model for any perfect fluid}
\label{sec:general}

The FRW cosmological models are characterized by the
scale factor $a(t)$ and have the following line element,
\begin{equation}  
\label{1}
ds^2 = - N^2(t) dt^2 + a^2(t)\left( \frac{dr^2}{1 - kr^2} + r^2 d\Omega^2
\right)\, ,
\end{equation}
where $d\Omega^2$ is the line element of the two-dimensional sphere with
unitary radius, $N(t)$ is the lapse function and $k$ gives the type of
constant curvature of the spatial sections. It may assume the values $k=-1, 1, 0$ 
and we are using the natural
unit system, where $c=G=1$. The matter content of the model is
represented by a perfect fluid with four-velocity $U^\mu = N(t)\delta^{\mu}_0$
in the comoving coordinate system used. The total energy-momentum tensor 
is given by,
\begin{equation}
T_{\mu \nu} = (\rho+p)U_{\mu}U_{\nu} + p g_{\mu \nu}\, ,  
\label{2}
\end{equation}
where $\rho$ and $p$ are the energy density and pressure of the fluid,
respectively. Here, we assume that $p = \alpha\rho$, where $\alpha$ is a constant
which defines the perfect fluid.

From the metric (\ref{1}) and the energy momentum tensor (\ref{2}), one may 
write the total Hamiltonian of the present model ($N {\mathcal{H}}$), where
${\mathcal{H}}$ is the superhamiltonian constraint.
It is given by \cite{germano1},

\begin{equation}
N {\mathcal{H}}= -\frac{P_{a}^2}{24} - 6ka^2 + a^{1-3\alpha}P_{T},  
\label{3}
\end{equation}
where $P_{a}$ and $P_{T}$ are the momenta canonically conjugated to $a$ and 
$T$, the latter being the canonical variable associated to the fluid \cite
{germano1}. Here, we are working in the conformal gauge, where $N = a$.

In order to introduce the noncommutativity in the model, we start considering,
initially, that the total Hamiltonian of the model has the same functional form 
as (\ref{3}). But now it is written in terms of noncommutative variables,
\begin{equation}
N_{nc} {\mathcal{H}}_{nc}= -\frac{P_{anc}^2}{24} - 6ka_{nc}^2 + a_{nc}^{1-3\alpha}P_{Tnc},  
\label{3,5}
\end{equation}
Then, we propose that the noncommutative variables of the model 
$\{a_{nc}, P_{anc}, T_{nc}, P_{Tnc}\}$ satisfy the following deformed Poisson brackets (PBs):
\ba
\label{4}
\left\{a_{nc},T_{nc}\right\}=\left\{P_{anc},P_{Tnc}\right\}=0,\\
\label{4.1}
\left\{a_{nc},P_{anc}\right\}=\left\{T_{nc},P_{Tnc}\right\}=1,\\
\label{4.2}
\left\{a_{nc},P_{Tnc}\right\}=\left\{T_{nc},P_{anc}\right\}=\gamma,
\ea 
in which $\gamma$ is the NC parameter. It is important to notice that this is
not the only possible deformed PBs one may propose, for the present model. 

As mentioned, above, in Ref. \cite{gil} the authors considered a very similar classical, noncommutative, FRW model coupled to a perfect fluid, 
in the presence of a cosmological constant. The only differences between our NC model and the NC model in Ref. \cite{gil} are the choices of deformed PBs and the 
presence of a cosmological constant in their model. In Section \ref{sec:comparison}, we shall make a detailed comparison between both models.
In their choice of deformed PBs, they made the two PBs in Eq. (\ref{4}) different from zero, instead of the two PBs in Eq. (\ref{4.2}). This can
be better seen in Eq. (\ref{35}). Therefore, since one of our motivations is investigating possible differences among different 
deformed PBs choices the only possibility, that does not include any of the PBs in Eq. (\ref{4}), was to make the two PBs in Eq. (\ref{4.2}) different from zero. For
simplicity we make them equal to the same NC parameter.

We would like to describe those models in terms of usual commutative variables, which satisfy the usual PBs.
Because it is simpler to deal with that kind of variables.
Following the literature of noncommutative theories it is possible to achieve that by introducing a set of 
coordinate transformations from the noncommutative variables to new commutative ones. Those type of 
transformations were first introduced in Refs. \cite{susskind} and sometimes are called Bopp shift \cite{zachos}.
Due to our choice of deformed PBs (\ref{4.2}), the more general transformations, to first order in $\gamma$, leading
from the NC variables to new commutative ones, are given by,
\ba
\label{5}
a_{nc}\rightarrow a_c + \frac{\gamma}{2}T_c,\nonumber\\
P_{anc}\rightarrow P_{ac} + \frac{\gamma}{2}P_{Tc},\nonumber\\
T_{nc}\rightarrow T_c + \frac{\gamma}{2}a_c,\\
P_{Tnc}\rightarrow P_{Tc} + \frac{\gamma}{2}P_{ac},\nonumber
\ea
where the commutative variables have $c$ labels. It is important to notice that
if we introduce the noncommutative variables Eq. (\ref{5}), in the deformed PBs Eq. (\ref{4}-\ref{4.2}) 
and use the usual PBs among the commutative variables, they are satisfied to first 
order in $\gamma$. Another important motivation to use those commutative variables, is that, the metric for those models
may be written in terms of them as,
\begin{eqnarray}  
\label{5,5}
ds^2 & = & - \left(a_c(t) + \frac{\gamma}{2}T_c(t)\right)^2 dt^2 \nonumber\\
& + & \left(a_c(t) + \frac{\gamma}{2}T_c(t)\right)^2\left( \frac{dr^2}{1 - kr^2} + r^2 d\Omega^2\right)\, .
\end{eqnarray}
For $\gamma = 0$, this metric reduces to Eq. (\ref{1}), in the gauge $N = a$. Observing the metric Eq. (\ref{5,5}), we notice that the dynamics of
world lines separations between two different times is given by the NC scale factor,
\begin{equation}
\label{5,55}
a_{nc}(t) = a_c(t) + \frac{\gamma}{2}T_c(t).
\end{equation}
Therefore, in our study of the dynamics of the models described by the metric Eq. ({\ref{5,5}), we must compute the NC scale factor given by Eq. (\ref{5,55}).
Since, all quantities in that metric Eq. ($\ref{5,5})$ are
commutative, we can treat those models using the usual general relativity methods. In particular, if we write the conservation equation for the
fluid stress-energy tensor Eq. (\ref{2}), for the metric Eq. (\ref{5,5}), we obtain the following relationship between the fluid density and
the NC scale factor,
\be
\label{5,6}
\rho(t) = \bar{C} \left(a_c(t) + \frac{\gamma}{2}T_c(t)\right)^{-3(\alpha + 1)},
\ee
where $\bar{C}$ is a positive constant. In terms of the commutative variables Eq. (\ref{5}), we have two equivalent ways to write the equations 
that describe the dynamics of the models. In the first one, we write the Einstein's equation for the metric Eq. (\ref{5,5}), use the expression 
for $\rho(t)$ Eq. (\ref{5,6}) and the equation of state for the fluid. In the second way, we introduce the transformations Eq. (\ref{5}) in the 
total Hamiltonian Eq. (\ref{3,5}) and compute the Hamilton's equations for the commutative variables. Since both ways are entirely equivalent, we 
shall use the second way.

We start rewriting the total Hamiltonian $N_{nc} {\mathcal{H}}_{nc}$ Eq. (\ref{3,5}), in terms of the commutative variables Eq. (\ref{5}),
\begin{eqnarray}
\label{6}
N_{nc} {\mathcal{H}}_{nc} & = & -\frac{1}{24} \left( P_{ac} + \frac{\gamma}{2}P_{Tc} \right)^2 
- 6k \left( a_c + \frac{\gamma}{2}T_c \right)^2\nonumber\\
& + & \left(a_c + \frac{\gamma}{2}T_c\right)^{1-3\alpha} \left(P_{Tc} + \frac{\gamma}{2}P_{ac}\right),  
\end{eqnarray}

The Hamilton's equations of motion, obtained using the total Hamiltonian Eq. (\ref{6}) and the
usual PBs among the commutative variables, are,
\ba
\dot a_c &=& \left\{a_c, N_{nc}{\cal{H}}_{nc}\right\} = -\frac{1}{12}\left(P_{ac} + \frac{\gamma}{2}P_{Tc}\right)\nonumber\\ 
&+& \frac{\gamma}{2}\left(a_c + \frac{\gamma}{2}T_c\right)^{1-3\alpha},\label{7}\\
\dot P_{ac} &=& \left\{P_{ac}, N_{nc}{\cal{H}}_{nc}\right\} = 12k\left( a_c + \frac{\gamma}{2}T_c \right)\nonumber\\
&-& (1 - 3\alpha)\left(a_c + \frac{\gamma}{2}T_c\right)^{-3\alpha}\left(P_{Tc} + \frac{\gamma}{2}P_{ac}\right),\label{8}\\
\dot T_c &=& \left\{T_c, N_{nc}{\cal{H}}_{nc}\right\} = -\frac{\gamma}{24}\left( P_{ac} + \frac{\gamma}{2}P_{Tc} \right) \nonumber\\
&+& \left(a_c + \frac{\gamma}{2}T_c\right)^{1-3\alpha},\label{9}\\
\dot P_{Tc} &=& \left\{P_{Tc}, N_{nc}{\cal{H}}_{nc}\right\} = 6\gamma k\left( a_c + \frac{\gamma}{2}T_c\right)\nonumber\\
&-& (1 - 3\alpha)\frac{\gamma}{2}\left(a_c + \frac{\gamma}{2}T_c\right)^{-3\alpha}\left(P_{Tc} + \frac{\gamma}{2}P_{ac}\right)
\label{10}
\ea
Now, we would like to find the NC scale factor behavior (\ref{5,55}). In the general situation, for generic $\alpha$ and $k$, the best we can
do is writing, from Eqs. (\ref{7})-(\ref{10}), a system of two coupled differential equations involving $a_c(t)$, $T_c(t)$
and their time derivatives. This is done in the following way. Combining Eqs. (\ref{8}) and (\ref{10}), we obtain the following
relationship between $P_{Tc}$ and $P_{ac}$,
\be
\label{11}
P_{Tc} = \frac{\gamma}{2}P_{ac} + C,
\ee
where $C$ is an integration constant. Physically, for the commutative case ($\gamma=0$), $C$ represents the fluid energy, which means 
that it is positive. Then, using Eqs. (\ref{7}) and (\ref{9}), we find, to first order in $\gamma$, the following equation
expressing $P_{ac}$ in terms of time derivatives of $a_c$ and $T_c$,
\be
\label{12}
P_{ac} = -12\dot a_c + 6\gamma\dot T_c - \frac{\gamma}{2}C.
\ee
Finally, we introduce the values of $\dot P_{ac}$ Eq. (\ref{8}), $\dot T_c$ Eq. (\ref{9}), $\dot P_{Tc}$ Eq. (\ref{10}), 
$P_{Tc}$ Eq. (\ref{11}) and $P_{ac}$ Eq. (\ref{12}), in the time derivative of Eq. (\ref{7}) and in Eq. (\ref{9}). It gives, to first
order in $\gamma$, the following system of coupled differential equation for $a_c$ and $T_c$,
\ba
\ddot a_c(t) &=& -k\left(a_c(t) + \frac{\gamma}{2}T_c(t)\right) \nonumber\\
&-& (1 - 3\alpha)\left(\frac{\gamma}{2}\dot a_c(t) - \frac{C}{12}\right)\left(a_c(t) +
\frac{\gamma}{2}T_c(t)\right)^{-3\alpha}\label{13}\\
\dot T_c(t) &=& \frac{\gamma}{2}\dot a_c(t) + \left(a_c(t) + \frac{\gamma}{2}T_c(t)\right)^{1-3\alpha}.\label{14}
\ea
All the information about the noncommutativity is encoded in the parameter $\gamma$. If we set it to zero we recover the usual
commutative model in the gauge $N=a$. In particular, equation (\ref{13}) decouples and we may solve it to obtain the scale factor
dynamics. In order to solve those equations and compute $a_{nc}$ Eq. (\ref{5,55}), we shall have to furnish initial conditions for $a_c(t)$, $\dot a_c(t)$ and $T_c(t)$. 
%We shall
%restrict our attention for the cases where $T_c(t=0) = 0$, because we want that initially the quantities Eq. (\ref{5,5}) and 
%$a_{c,\gamma=0}$, are equal. The last quantity is obtained from $a_c$ by setting $\gamma \to 0$. 
%We shall consider, here,
%models starting from a {\it Big Bang} singularity and with an initial positive $\dot a_c(t)$, therefore: $a_c(t=0) = 0$ and 
%$\dot a_c(t=0) = \dot a_{c0} > 0$. As a matter of simplicity and because no qualitative differences appeared for the models studied
%here, when we varied the value of $\dot a_{c0}$, we shall fix the value of this quantity. Then, in what follows we shall use the
%initial conditions,
%\be
%\label{17}
%a_c(t=0) = 0,\qquad \dot a_c(t=0) = 1.
%\ee
Unfortunately, we cannot find algebraic solutions for $a_c(t)$ and $T_c(t)$, from the system Eqs. (\ref{13})-(\ref{14}), for generic 
values of $\alpha$ and $\gamma$. On the other hand, we may solve them for certain types of perfect fluids. Therefore, in what follows,
we shall investigate two important cases, where we can find algebraic solutions for $a_c(t)$ and $T_c(t)$, the 
radiation perfect fluid and the dust perfect fluid.

\section{The noncommutative model for a radiation perfect fluid}
\label{sec:radiation}

Let us consider, initially, the case of a radiation perfect fluid. Therefore, we assume that $p = \rho/3$, which is the equation of
state for radiation. This choice may be considered as a first approximation to treat the matter content of the early Universe and it 
was made as a matter of simplicity. It is clear that a more complete treatment should describe the radiation, present in the primordial 
Universe, in terms of the electromagnetic field.

If we introduce $\alpha=1/3$ in Eq. (\ref{14}), we may integrate the resulting equation, in the $t$ variable, to obtain $T_c$ as the 
following function of $a_c$.
\be
\label{15}
T_c(t) = \frac{\gamma}{2} ( a_c(t) - a_0 ) + t + T_0.
\ee
Where $a_0$ and $T_0$ are, respectively, the initial values ($t = 0$) of $a_c(t)$ and $T_c(t)$. If we introduce $\alpha=1/3$ and the 
value of $T(t)$ Eq. (\ref{15}), in Eq. (\ref{13}), we obtain, to first order in $\gamma$, the following differential equation for $a_c(t)$,
\be
\label{16}
\ddot{a}_c(t) + k a_c(t) + \frac{\gamma}{2} k (t + T_0) = 0,
\ee
If we set $\gamma=0$, in Eq. (\ref{16}), the commutative FRW cosmological model is restored. In what follows, we will
solve Eq.(\ref{16}), analytically, for $k = 0, \pm 1$. Then, we will compute the NC scale factor Eq. (\ref{5,55}), with 
the aid of $a_c(t)$ Eq. (\ref{16}) and $T_c(t)$ Eq. (\ref{15}). Finally, we will explore some features of those solutions, 
for several values of $\gamma$.

\subsection{The case $k = 0$}
\label{subsec:radiationk0}

For spacetimes with flat spatial sections, $k=0$, Eq. (\ref{16}), reduces to the commutative case. If we solve this equation, 
we obtain the following scale factor expression,
\be
\label{16,5}
a_c(t) =  v_0 t + a_0,
\ee
where $v_0$ is the initial value ($t = 0$) of $\dot{a}_c(t)$. This is exactly the scale factor in the usual commutative model, 
which we will write $a_{c,\gamma=0}(t)$. As mentioned, above, the NC scale factor Eq. (\ref{5,55}) is the physical one associated 
with the metric Eq. (\ref{5,5}). In order to compute it, we start evaluating $T_c(t)$ Eq. (\ref{15}),
with the aid of $a_c(t)$ Eq. (\ref{16,5}). Finally, we obtain the NC scale factor Eq. (\ref{5,55}), to first order in $\gamma$,
with the aid of $T_c(t)$ and $a_c(t)$ Eq. (\ref{16,5}),
\be
\label{16,6}
a_{nc}(t) = \left(v_0 + \frac{\gamma}{2}\right) t + a_0 + T_0.
\ee
In order to compare the NC scale factor Eq. (\ref{16,6}) with the usual commutative one $a_{c,\gamma=0}(t)$, which in this 
case is given by Eq. (\ref{16,5}), let us set $T_0 = 0$ in Eq. (\ref{16,6}). It is done in order to assure that both of them
have the same initial value $a_0$. Then, we notice that both scale factors are linear functions of $t$, with the same
$a$-intercept and different slopes. For $v_0>0$, if $\gamma >0$, the NC scale factor increases more rapidly than $a_{c,\gamma=0}(t)$ 
Eq. (\ref{16,5}). On the other hand, if $\gamma < 0$ and $v_0>|\gamma/2|$, the NC scale factor increases slower than
$a_{c,\gamma=0}(t)$. Finally, if $\gamma < 0$ and $v_0<|\gamma/2|$, the NC scale factor decreases from $a_0$ until the
{\it Big Crunch}.

\subsection{The case $k = 1$}
\label{subsec:radiationk1}

For spacetimes with constant positive spatial sections, $k=1$, Eq. (\ref{16}), describes a driven harmonic oscillator, 
under the driving force $-(\gamma/2) (t + T_0)$. If we solve this equation, we obtain the following scale factor expression,
\be
\label{18}
a_c(t) = \left(v_0 + \frac{\gamma}{2}\right)\sin(t) + \left(a_0 + \frac{\gamma}{2} T_0\right) \cos(t) - 
\frac{\gamma}{2} (t + T_0).
\ee
Now, in order to compute the NC scale factor Eq. (\ref{5,55}), we start evaluating $T_c(t)$ Eq. (\ref{15}). 
With the aid of $a_c(t)$ Eq. (\ref{18}), it is given, to first order in $\gamma$, by,
\be
\label{18,5}
T_c(t) = \frac{\gamma}{2}\left(v_0\sin(t) + a_0\cos(t) - a_0\right) + t + T_0.
\ee
Finally, using $a_c(t)$ Eq. (\ref{18}) and $T_c(t)$ Eq. (\ref{18,5}), the NC scale factor Eq. (\ref{5,55}) is given, to 
first order in $\gamma$, by,
\be
\label{18,6}
a_{nc}(t) = \left(v_0 + \frac{\gamma}{2}\right)\sin(t) + \left(a_0 + \frac{\gamma}{2} T_0\right) \cos(t).
\ee
Now, we would like to compare the NC scale factor Eq. (\ref{18,6}) with the usual commutative one $a_{c,\gamma=0}(t)$.
We shall set $T_0=0$ in $a(t)$ Eq. (\ref{18,6}). In order to obtain $a_{c,\gamma=0}(t)$, we set $\gamma=0$ in Eqs. (\ref{18}) or (\ref{18,6}). 
It gives,
\be
\label{19}
a_{c,\gamma=0}(t) = v_0 \sin(t) + a_0 \cos(t).
\ee

The commutative solution Eq. (\ref{19}) describes an universe that starts expanding from $a_0$ at $t=0$, 
expands to a maximum size and then contracts back to a {\it Big Crunch} singularity, in a finite time. On the other
hand, the NC solution Eq. (\ref{18,6}), may describe different scale factor behaviors. We studied that
solution for several different values of $\gamma$ and reached the following conclusions, for any values of $a_0$ and $v_0$. For positive $\gamma$,
the NC solutions have the same general behavior than the commutative solution, except that their 
maximum sizes are bigger than the commutative one and they take greater times, than the commutative one, to return to $a=0$. 
If we increase the absolute value of $\gamma$, that behavior becomes even more pronounced. Therefore, we conclude that when 
$\gamma$ is positive, the resulting effect upon the scale factor dynamics is the appearance of an additional repulsive 
force, compared to the commutative case. An example of this case is shown in Figure 1. For negative $\gamma$, 
the NC solutions have the same general behavior than the commutative solution. The differences are that their maximum 
sizes are smaller than the commutative one and they take shorter times, than the commutative solution, to return to $a=0$. 
If we increase the absolute values of $\gamma$, that behavior becomes even more pronounced. Therefore, we conclude that when
$\gamma$ is negative, the resulting effect upon the scale factor dynamics is the appearance of an additional attractive force,
compared to the commutative case. An example of this case is shown in Figure 2.

\begin{figure}
\includegraphics[width=6cm,height=6cm,angle=0]{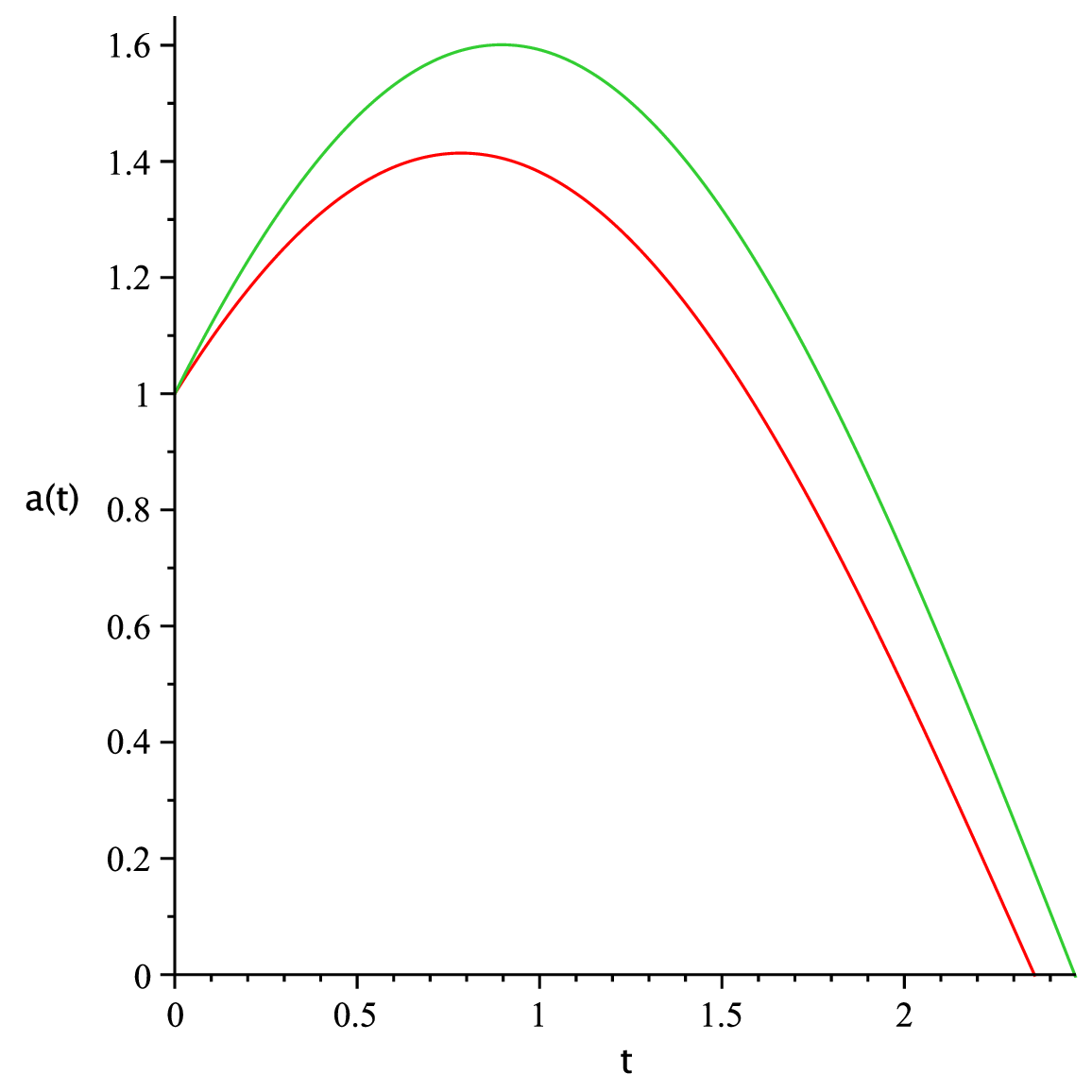}
\caption{{\protect\footnotesize {Scale factor behavior for both commutative and noncommutative cases, when $k=1$. Here, we consider
$\gamma=0.5$, $a_0=1$ and $v_0=1$. The $a(t)$ for the noncommutative case is the upper curve.}}}
\label{fig1}
\end{figure}

\begin{figure}
\includegraphics[width=6cm,height=6cm,angle=0]{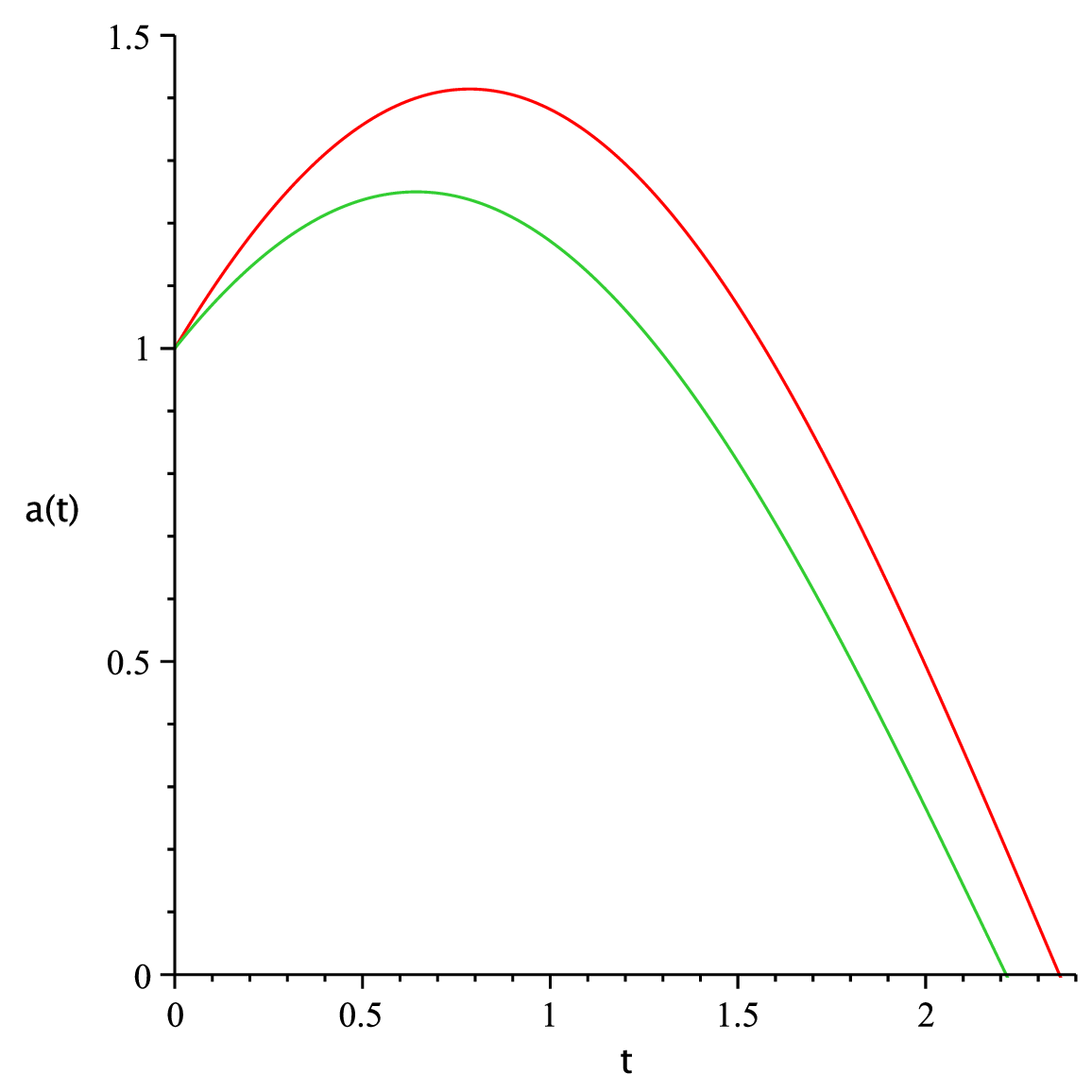}
\caption{{\protect\footnotesize {Scale factor behavior for both commutative and noncommutative cases, when $k=1$. Here, we consider
$\gamma=-0.5$, $a_0=1$ and $v_0=1$. The $a(t)$ for the commutative case is the upper curve.}}}
\label{fig2}
\end{figure}

%\begin{figure}
%\includegraphics[width=6cm,height=6cm,angle=0]{Figure3.eps}
%\caption{{\protect\footnotesize {Scale factor behavior for both commutative and noncommutative cases, when $k=1$. Here, we consider
%$\gamma=-0.25$ and $C_1=3$. The $a(t)$ for the noncommutative case is the one which grows with time.}}}
%\label{fig3}
%\end{figure}

\subsection{The case $k = -1$}
\label{subsec:radiationk-1}

Solving Eq. (\ref{16}) for spacetimes with constant negative spatial sections, $k=-1$,
we obtain the following scale factor expression,
\be
\label{20}
a_c(t) = \left(v_0 + \frac{\gamma}{2}\right)\sinh(t) + \left(a_0 + \frac{\gamma}{2} T_0\right) \cosh(t) - \frac{\gamma}{2} (t + T_0).
\ee
Now, in order to compute the NC scale factor Eq. (\ref{5,55}), we start evaluating $T_c(t)$ Eq. (\ref{15}). 
With the aid of $a_c(t)$ Eq. (\ref{20}), it is given, to first order in $\gamma$, by,
\be
\label{20,5}
T_c(t) = \frac{\gamma}{2}\left(v_0\sinh(t) + a_0\cosh(t) - a_0\right) + t + T_0.
\ee
Finally, using $a_c(t)$ Eq. (\ref{20}) and $T_c(t)$ Eq. (\ref{20,5}), the NC scale factor Eq. (\ref{5,55}) is given, to 
first order in $\gamma$, by,
\be
\label{20,6}
a_{nc}(t) = \left(v_0 + \frac{\gamma}{2}\right)\sinh(t) + \left(a_0 + \frac{\gamma}{2} T_0\right) \cosh(t).
\ee
Now, we would like to compare the NC scale factor Eq. (\ref{20,6}) with the usual commutative one $a_{c,\gamma=0}(t)$.
We shall set $T_0=0$ in $a(t)$ Eq. (\ref{20,6}). In order to obtain $a_{c,\gamma=0}(t)$, we set $\gamma=0$ in Eqs. (\ref{20}) or (\ref{20,6}). 
\be
\label{21}
a_{c,\gamma=0}(t) = v_0 \sinh(t) + a_0 \cosh(t).
\ee

The commutative solution Eq. (\ref{21}) describes an universe that starts from $a_0$ at $t=0$ and 
expands exponentially to an infinity size when the time goes to infinity. On the other hand, the noncommutative solution 
Eq. (\ref{20,6}), may describe different scale factor behaviors. We studied that solution for several different values of 
$\gamma$ and reached the following conclusions, for any value of $a_0$. For positive $\gamma$ and any value of $v_0$, the noncommutative solutions 
have the same general behavior than the commutative solution, except that they expand in a faster rate than that solution.
If we increase the absolute values of $\gamma$, that behavior becomes even more pronounced. Therefore, we conclude that when $\gamma$ 
is positive, the resulting effect upon the scale factor dynamics is the appearance of an additional repulsive 
force, compared to the commutative case. An example of this case is shown in Figure 3. For negative $\gamma$ and $v_0 > |\gamma/2|$,
the NC scale factor has the same general behavior than the commutative solution, except that it expands in a slower rate than that solution.
An example of this case is shown in Figure 4.
For negative $\gamma$ and $v_0 < |\gamma/2|$, the NC scale factor starts contracting from $a_0$ until it reaches a minimum value. Then, it
expands exponentially to infinity in a slower rate than the commutative solution. An example of this case is shown in Figure 5.
If we increase the absolute values of $\gamma$, that behavior becomes even more pronounced. Therefore, we conclude that when $\gamma$ 
is negative, the resulting effect upon the scale factor dynamics is the appearance of an additional attractive force, compared to 
the commutative case.

\begin{figure}
\includegraphics[width=6cm,height=6cm,angle=0]{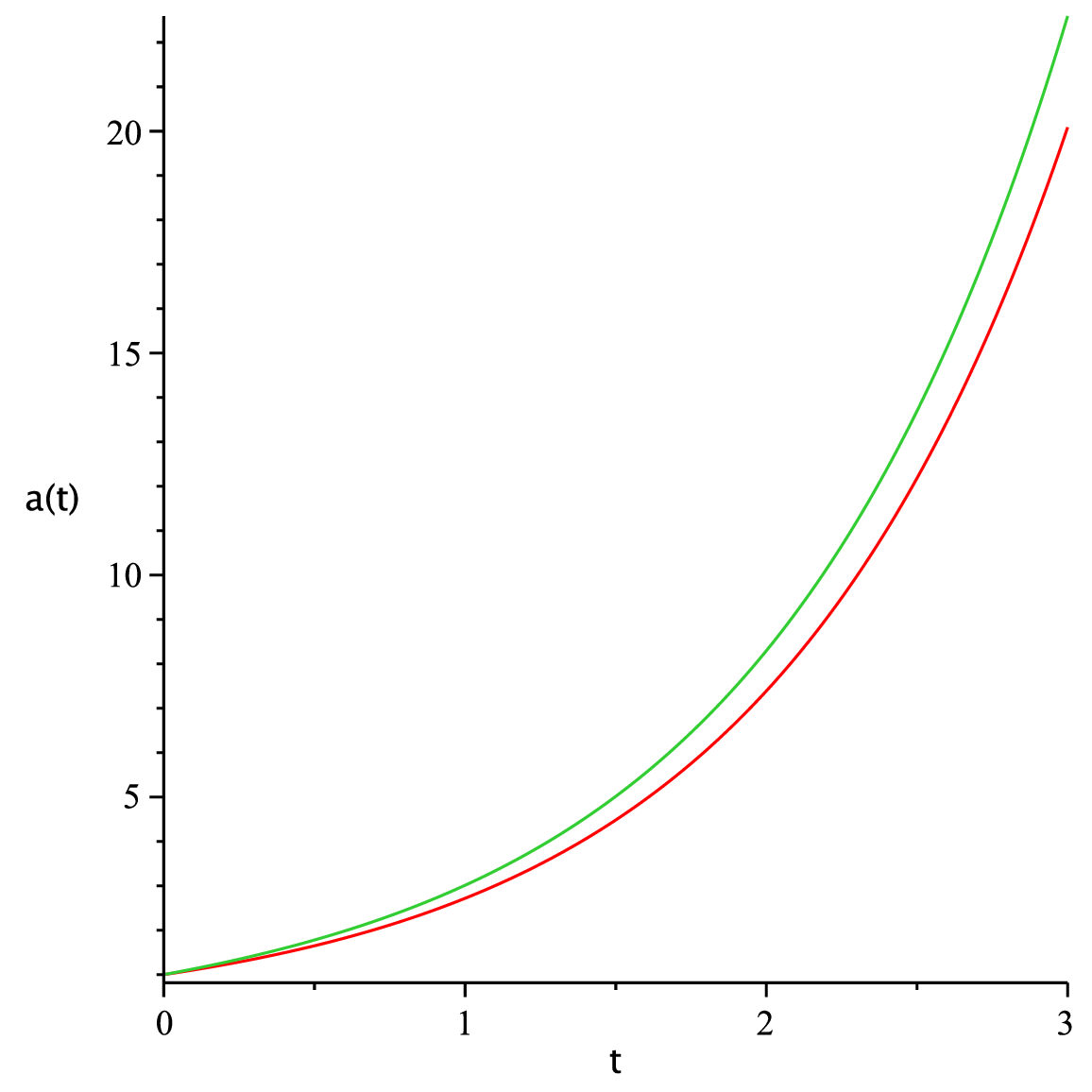}
\caption{{\protect\footnotesize {Scale factor behavior for both commutative and noncommutative cases, when $k=-1$. Here, we consider
$\gamma=0.5$, $a_0=1$ and $v_0=1$. The $a(t)$ for the noncommutative case is the upper curve.}}}
\label{fig3}
\end{figure}

\begin{figure}
\includegraphics[width=6cm,height=6cm,angle=0]{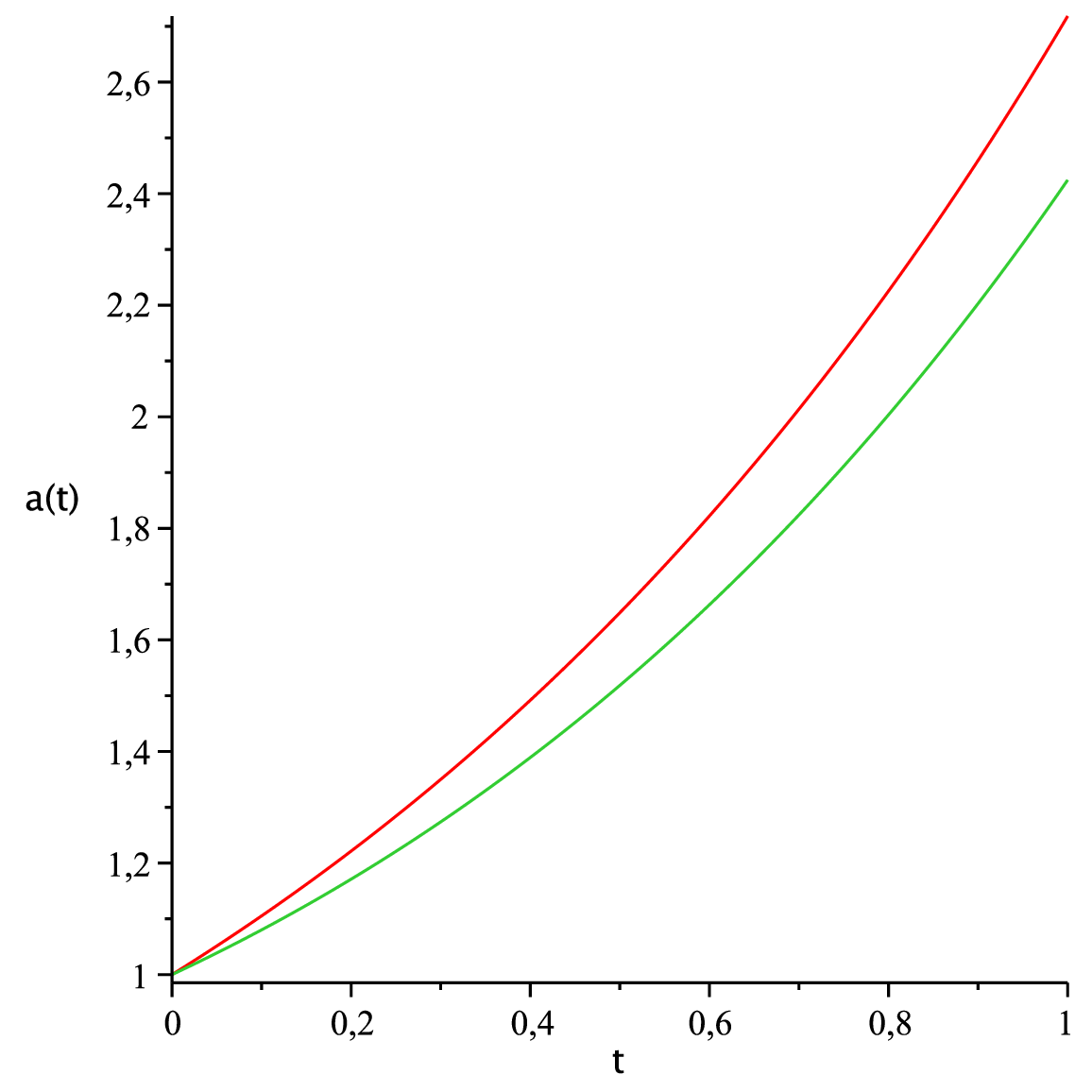}
\caption{{\protect\footnotesize {Scale factor behavior for both commutative and noncommutative cases, when $k=-1$. Here, we consider
$\gamma=-0.5$, $a_0=1$ and $v_0=1$. The $a(t)$ for the commutative case is the upper curve.}}}
\label{fig4}
\end{figure}

\begin{figure}
\includegraphics[width=6cm,height=6cm,angle=0]{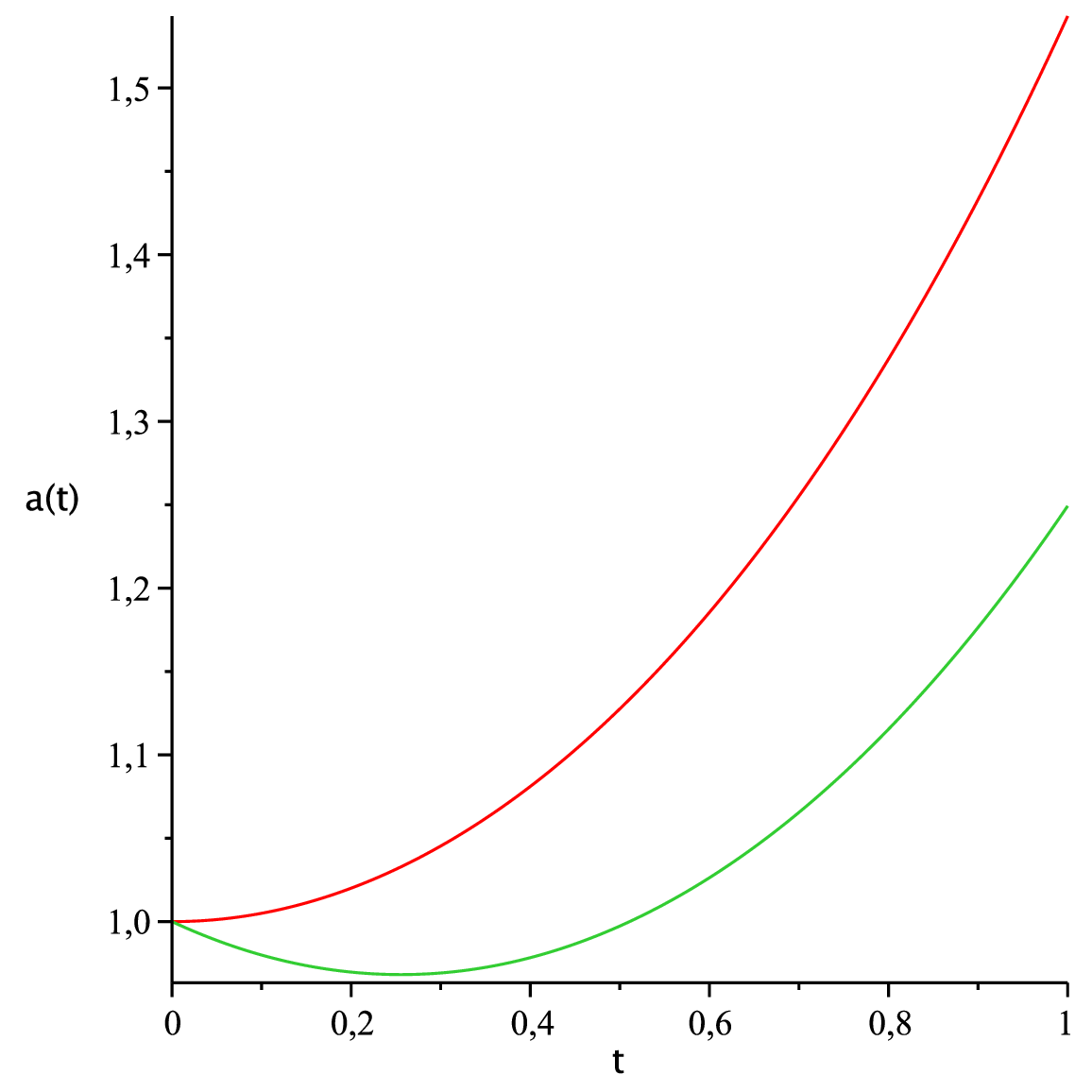}
\caption{{\protect\footnotesize {Scale factor behavior for both commutative and noncommutative cases, when $k=-1$. Here, we consider
$\gamma=-0.5$, $a_0=1$ and $v_0=0$. The $a(t)$ for the commutative case is the upper curve.}}}
\label{fig4}
\end{figure}

%\begin{figure}
%\includegraphics[width=6cm,height=6cm,angle=0]{Figure6.eps}
%\caption{{\protect\footnotesize {Scale factor behavior for both commutative and noncommutative cases, when $k=-1$. Here, we consider
%$\gamma=0.1$ and $C_1=-40$. The $a(t)$ for the noncommutative case is the one that initially expands and then contracts.}}}
%\label{fig6}
%\end{figure}

\section{The noncommutative model for a dust perfect fluid}
\label{sec:dust}

Let us consider, now, the case of a dust perfect fluid. Therefore, we assume that $p = 0$, which is the equation of
state for dust. This choice may be considered as a first approximation to treat the matter content of the present Universe which is
dominated by matter that interacts weakly.

If we introduce $\alpha=0$ in Eqs. (\ref{13})-(\ref{14}), we obtain the new system of equations for $a_c$ and $T_c$,
\ba
\ddot a_c(t) &=& -k\left(a_c(t) + \frac{\gamma}{2}T_c(t)\right)
-\frac{\gamma}{2}\dot a_c(t) + \frac{C}{12},\label{22}\\
\dot T_c(t) &=& \frac{\gamma}{2}\dot a_c(t) + a_c(t) + \frac{\gamma}{2}T_c(t).\label{23}
\ea
%In order to solve this system, we have to consider two different cases: $k=0$ and $k\neq 0$. When $k=0$, we 
%may integrate Eq. (\ref{22}), in order to find $a(t)$. When $k\neq 0$, we, initially, have to multiply Eq. (\ref{23}) by $k$ 
%and sum the resulting equation with Eq. (\ref{22}). It gives,
%\be
%\label{24}
%\ddot a_c(t) + k\dot T_c(t) = \frac{\gamma}{2}(k-1)\dot a_c(t) + \frac{C}{12}.
%\ee
%After integrating Eq. (\ref{24}), once in time, we take the value of $T_c(t)$ from the resulting equation and introduce it in
%Eq. (\ref{22}). To first order in $\gamma$, it gives rise to the following equation,
%\be
%\label{25}
%\ddot a_c(t) + ka_c(t) + \frac{\gamma}{2}\left(\frac{C}{12}t + C_1\right) - \frac{C}{12} = 0,
%\ee
%where $C_1$ is an integration constant, which may be positive or negative, just like in the radiation case Section \ref{sec:radiation}. 
Here, also, if we set $\gamma=0$ the commutative FRW cosmological model is restored. In what follows, 
we will solve the system Eqs. (\ref{22})-(\ref{23}) to find $a_c(t)$ and $T_c(t)$. Subsequently, we shall combine those solutions
to compute the NC scale factor Eq. (\ref{5,55}). Finally, we shall explore some features of $a_{nc}(t)$, for several values of
$\gamma$, $C$ and $k =0, \pm 1$. We shall, also, compare the present results with the ones of Section \ref{sec:radiation}, for a radiation
perfect fluid.

\subsection{The case $k = 0$}
\label{subsec:dustk0}

Setting $k=0$ in Eq. (\ref{22}), we obtain,
\be
\label{26}
\ddot a_c(t) = -\frac{\gamma}{2}\dot a_c(t) + \frac{C}{12}.
\ee
This equation describes the scale factor dynamics under the action of two forces: a positive constant one and a velocity dependent one. 
%If $\gamma$ is 
%positive the velocity dependent force acts as a resistance force and decelerates the motion, on the other hand if $\gamma$ is negative it helps 
%accelerating the motion. 
For $\gamma=0$, we obtain, from Eq. (\ref{26}), the scale factor equation for the commutative case.
The solution to that equation is given by,
\be
\label{27}
a_{c,\gamma=0}(t) = \frac{1}{24} C t^2 + v_0 t + a_0.
\ee
Now, integrating Eq. (\ref{26}), we obtain the following expression for $a_c(t)$,
\be
\label{28}
a_c(t) = - \frac{e^{- \gamma t/2} (6\gamma v_0 - C)}{3\gamma^2} + \frac{C t}{6\gamma} + \frac{3\gamma^2 a_0 + 6\gamma v_0 - C}{3\gamma^2}.
\ee
If we take the limit $\gamma \to 0$ of $a_c(t)$ Eq. (\ref{28}), we obtain the commutative solution $a_{c,\gamma=0}(t)$ Eq. (\ref{27}).
Now, in order to compute the NC scale factor Eq. (\ref{5,55}), we must start solving equation (\ref{23}) in order to find $T_c(t)$. 
With the aid of $a_c(t)$ Eq. (\ref{28}), it is given, to first order in $\gamma$, by,
\ba
\label{28,5}
&T_c(t)& = \exp(\gamma t/2)\left(T_0 + \frac{2a_0}{\gamma}\right)\nonumber\\
&+& \left(v_0 + \frac{2C}{3\gamma^3}\right)\sinh(\gamma t/2) + \left(\frac{C}{6\gamma} + \frac{4v_0}{\gamma^2}\right)\cosh(\gamma t/2)\nonumber\\ 
&-& \frac{2a_0}{\gamma} - \frac{C}{6\gamma} - \frac{4v_0}{\gamma^2} - \frac{Ct}{3\gamma^3}.
\ea
Finally, using $a_c(t)$ Eq. (\ref{28}) and $T_c(t)$ Eq. (\ref{28,5}), the NC scale factor Eq. (\ref{5,55}) is given, to 
first order in $\gamma$, by,
\ba
\label{28,6}
&a_{nc}(t)& = \frac{v_0}{2\gamma}(4 + \gamma^2)\sinh(\gamma t/2)\nonumber\\ 
&+& \frac{C}{12\gamma^2}(4 + \gamma^2)\cosh(\gamma t/2) - \frac{C}{12\gamma^2}(4 + \gamma^2)\nonumber\\
&+& \frac{1}{2}(\gamma T_0 + 2a_0)\exp(\gamma t/2).
\ea
Now, we would like to compare the NC scale factor Eq. (\ref{28,6}) with the usual commutative one $a_{c,\gamma=0}(t)$ Eq. (\ref{27}).
We shall set $T_0=0$ in $a_{nc}(t)$ Eq. (\ref{28,6}).

The commutative solution Eq. (\ref{27}) describes an universe that starts from $a_0$ at $t=0$ and 
expands, as a second degree polynomial in $t$, to an infinity size when $t$ goes to infinity. On the other hand, the noncommutative solution 
Eq. (\ref{28,6}), may describe different scale factor behaviors. We studied that solution for several different values of 
$\gamma$ and $C$ and reached the following conclusions. For 
positive $\gamma$, the noncommutative solutions have the same general behavior than the commutative solution, except that 
they expand in a faster rate than the commutative one. In fact, it expands as an exponential function of $t$. 
If we increase $\gamma$, that behavior becomes even more pronounced. 
If we increase $C$, both commutative and noncommutative solutions increase their expansion rates, but the noncommutative 
solutions still expand faster than the commutative one. Those results are valid for any values of $a_0$ and $v_0$. 
An example of this case is shown in Figure 6. For negative $\gamma$, the noncommutative solutions have the same general 
behavior than the commutative solution, except that they initially expand in a slower rate than the commutative one. 
Then, they increase their expansion rate and eventually overtake the commutative solution. From that moment onward the 
noncommutative solutions expand in a faster rate than the commutative one. An example of this case is shown in Figure 7. 
For sufficiently large values of $a_0$ the 
noncommutative solutions, initially, decrease until reach a minimum value. Then, they start increasing until overtake the
commutative solution. An example of this case is shown in Figure 8. For sufficiently large values of $v_0$ the noncommutative 
solutions always expand faster than the commutative one. An example of this case is shown in Figure 9. If we increase the modulus of $\gamma$, those 
behaviors becomes even more pronounced. If we increase $C$, both commutative and noncommutative solutions increase their 
expansion rates, but the previous behaviors still take place.
From Eq. (\ref{28,6}), we can see that for sufficiently large values of $t$, the NC scale factor has an exponential expansion.
Since, the dust perfect fluid represents the present matter dominated era of our Universe and $k=0$ is a good candidate to describe its spatial curvature, 
then, under those circumstances, our NC model may be considered a possible candidate to describe the present expansion of the Universe.

\begin{figure}
\includegraphics[width=6cm,height=6cm,angle=0]{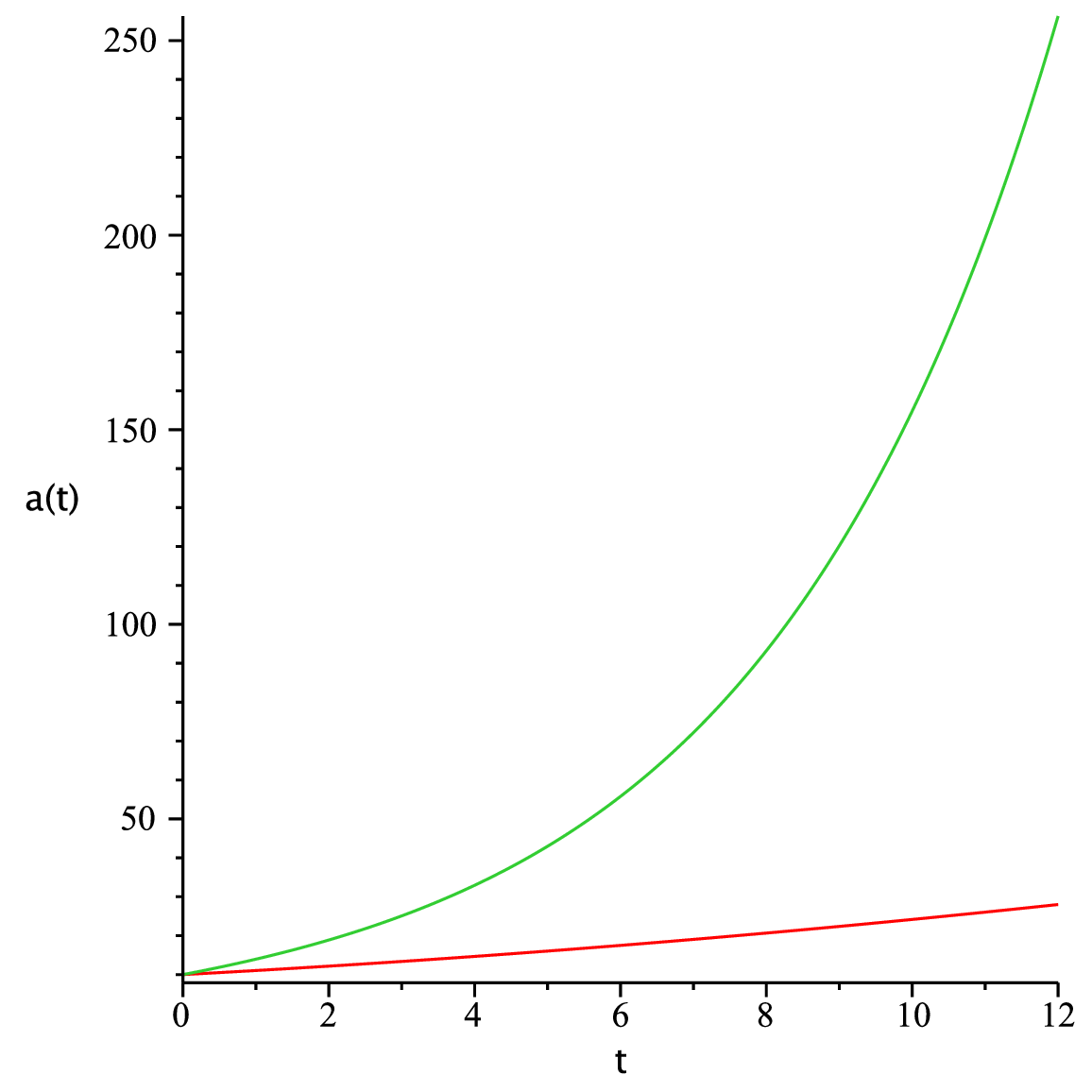}
\caption{{\protect\footnotesize {Scale factor behavior for both commutative and noncommutative cases, when $k=0$. Here, we consider
$\gamma=0.5$, $C=1$ $a_0=10$ and $v_0=1$. The $a(t)$ for the noncommutative case is the upper curve.}}}
\label{fig5}
\end{figure}

\begin{figure}
\includegraphics[width=6cm,height=6cm,angle=0]{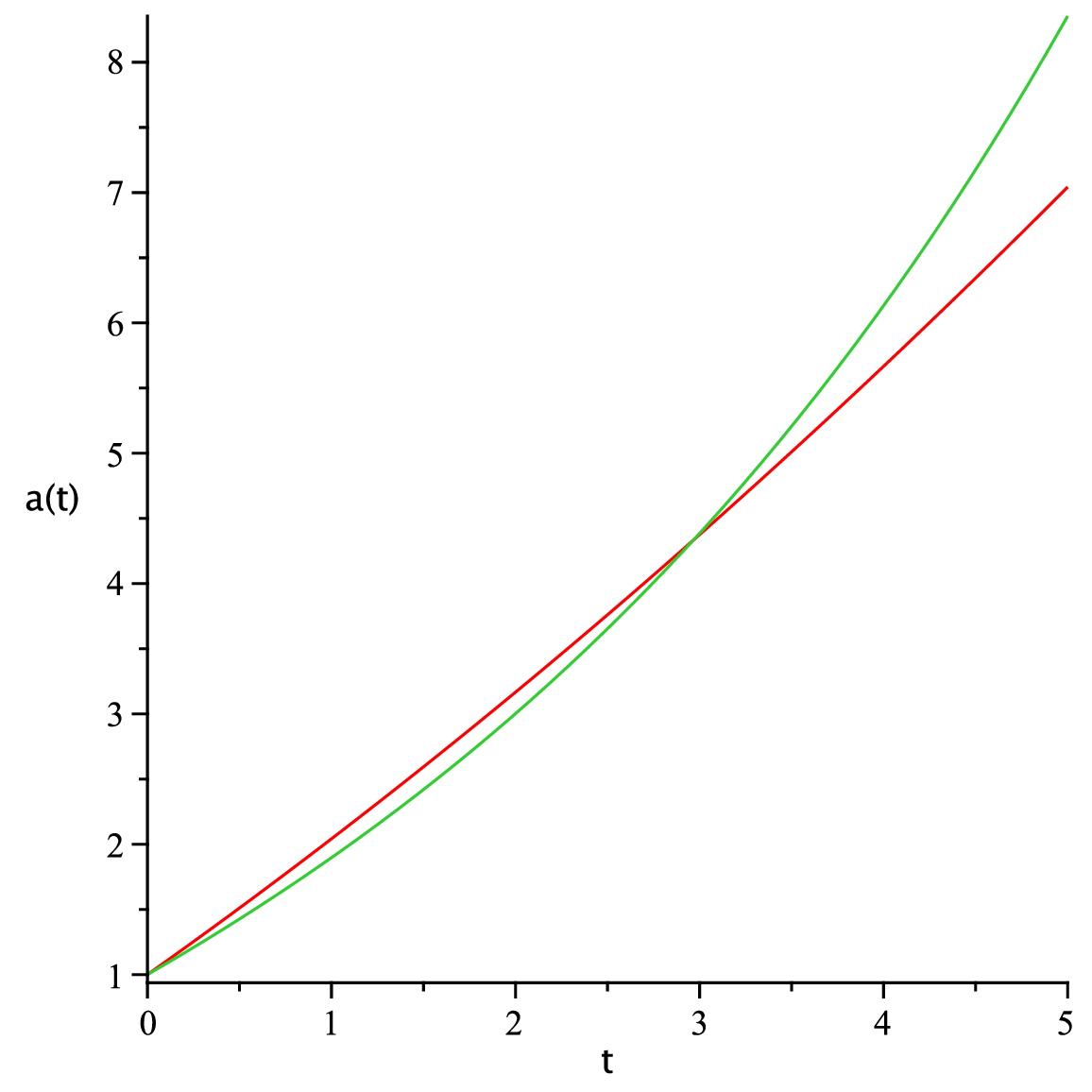}
\caption{{\protect\footnotesize {Scale factor behavior for both commutative and noncommutative cases, when $k=0$. Here, we consider
$\gamma=-0.5$, $C=1$ $a_0=1$ and $v_0=1$. The $a(t)$ for the noncommutative case is the one that expands asymptotically in a faster rate.}}}
\label{fig5.5}
\end{figure}

\begin{figure}
\includegraphics[width=6cm,height=6cm,angle=0]{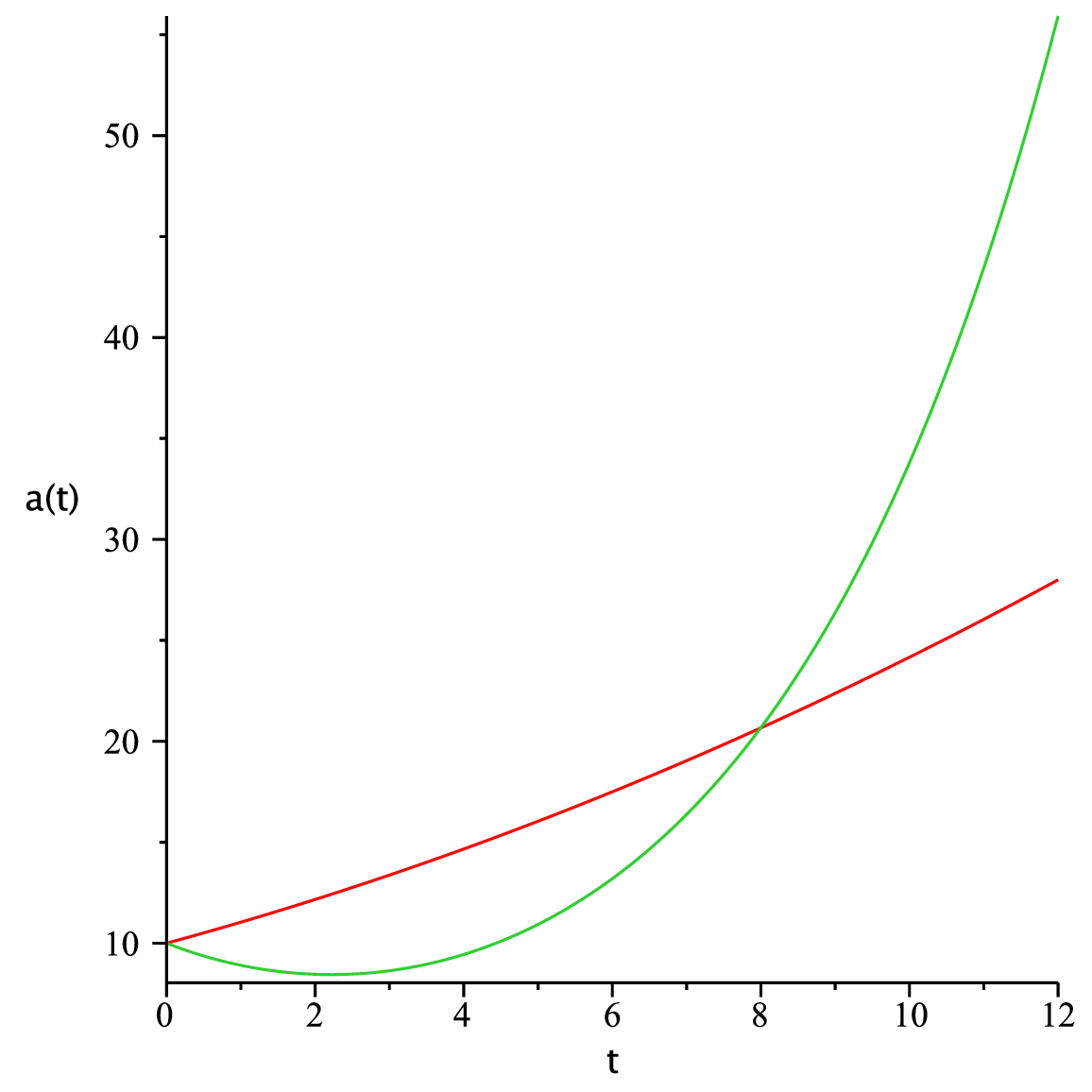}
\caption{{\protect\footnotesize {Scale factor behavior for both commutative and noncommutative cases, when $k=0$. Here, we consider
$\gamma=-0.5$, $C=1$ $a_0=10$ and $v_0=1$. The $a(t)$ for the commutative case is the one that is strictly increasing.}}}
\label{fig6}
\end{figure}

\begin{figure}
\includegraphics[width=6cm,height=6cm,angle=0]{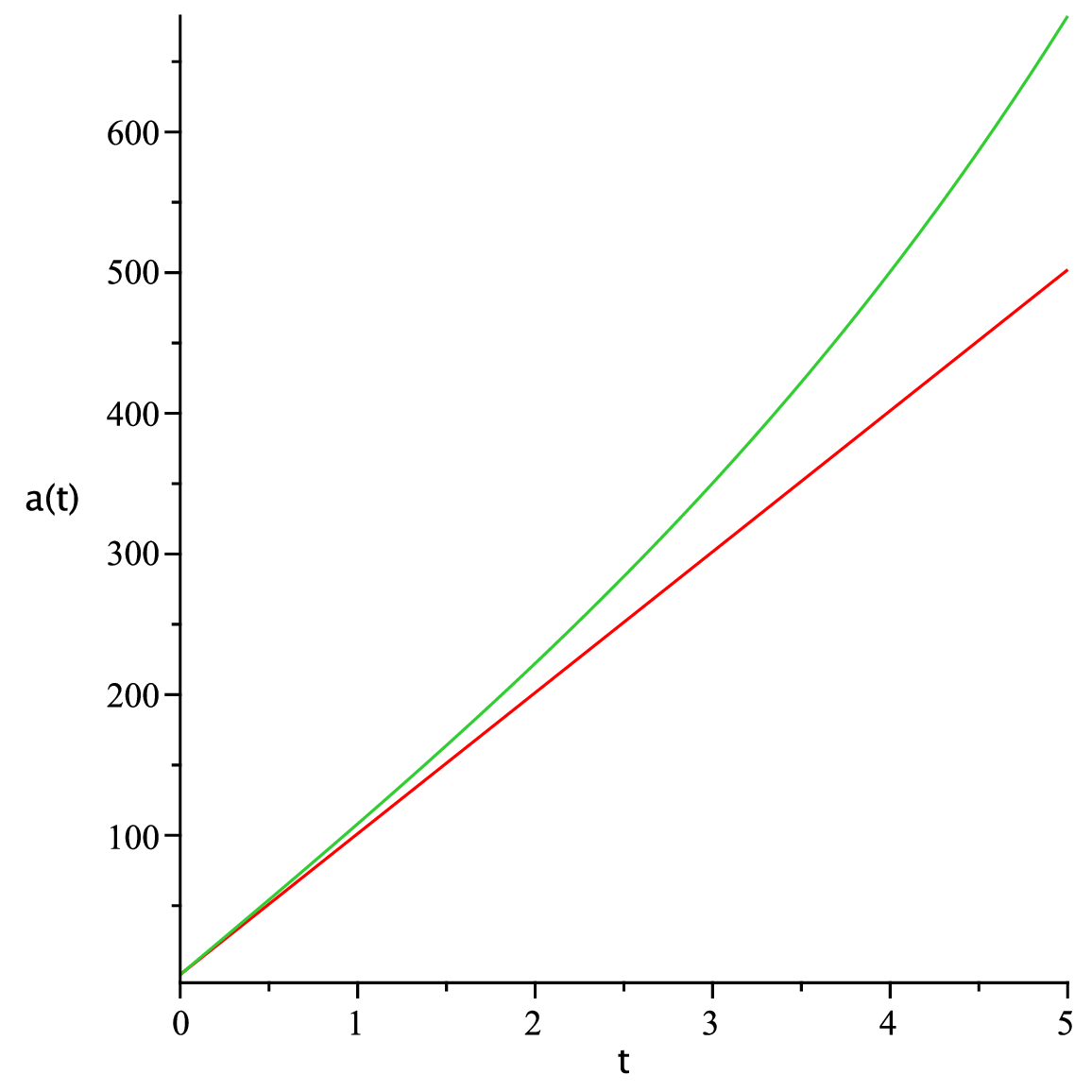}
\caption{{\protect\footnotesize {Scale factor behavior for both commutative and noncommutative cases, when $k=0$. Here, we consider
$\gamma=-0.5$, $C=1$ $a_0=1$ and $v_0=100$. The $a(t)$ for the noncommutative case is the upper curve.}}}
\label{fig6.5}
\end{figure}

\subsection{The case $k = 1$}
\label{subsec:dustk1}

If we set $k=1$ in the system Eqs. (\ref{22})-(\ref{23}), we may integrate it to find the following algebraic solutions for $a_c(t)$,
%In the same way as in the radiation case, for spacetimes with constant positive spatial sections, $k=1$, Eq. (\ref{25}) 
%is the equation of a driven harmonic oscillator, under the driving force $-(\gamma/2) (Ct/12 + C_1) + C/12$. We notice that 
%the driving force is different from the one in the radiation case. Solving this equation with initial conditions 
%Eq. (\ref{17}), we obtain the following scale factor expression,
\ba
\label{29}
a_c(t) &=& \left(a_0 + \frac{\gamma}{2}(v_0 + T_0) - \frac{C}{12}\right)\cos(t)\\\nonumber
&+& \left(v_0 + \frac{\gamma C}{24}\right)\sin(t) - \frac{\gamma}{2}\left(\frac{C}{12}t + v_0 + T_0\right) + \frac{C}{12},
\ea
and $T_c(t)$,
\ba
\label{29,5}
T_c(t) &=& \left(a_0 + \frac{\gamma}{2}(v_0 + T_0) - \frac{C}{12}\right)\sin(t)\\\nonumber
&-& \left(v_0 + \frac{\gamma C}{24}\right)\cos(t) + \frac{C}{12}t + v_0 + T_0 + \frac{\gamma C}{24}.
\ea
Now, we can compute $a_{nc}$ Eq. (\ref{5,55}), with the aid of $a_c(t)$ Eq. (\ref{29}) and $T_c(t)$ Eq. (\ref{29,5}).
It is given, to first order in $\gamma$, by,
\ba
\label{29,6}
a_{nc}(t) &=& \left(a_0 + \frac{\gamma}{2}T_0 - \frac{C}{12}\right)\cos(t)\nonumber\\
&+& \left(v_0 + \frac{\gamma}{2}a_0\right)\sin(t) + \frac{C}{12}.
\ea
If we set $\gamma=0$ in Eqs. (\ref{29}) or (\ref{29,6}), we obtain the scale factor in the usual commutative model,
\be
\label{30}
a_{c,\gamma=0}(t) = \left(a_0 - \frac{C}{12}\right)\cos(t) + v_0\sin(t) + \frac{C}{12}.
\ee

Now, we would like to compare the NC scale factor Eq. (\ref{29,6}) with the usual commutative one $a_{c,\gamma=0}(t)$ Eq. (\ref{30}).
We shall set $T_0=0$ in $a_{nc}(t)$ Eq. (\ref{29,6}). 
%The results described bellow will be valid for any values of $a_0$. The different
%values of $C$ in the domains of that constant described bellow will depend on the value of $v_0$. The greater $v_0$ the greater is the value of $C$ in each 

The commutative solution Eq. (\ref{30}) describes an universe that starts expanding from $a_0$ at $t=0$, expands to a maximum size and then
contracts. For small values of $C$ it will contract to a {\it Big Crunch} singularity. If we increase the value of $C$, both the maximum size and
the time interval between $a_0$ and the {\it Big Crunch} increase. This behavior is similar to the radiation case Subsection \ref{subsec:radiationk1}. 
For sufficiently large values of $C$ the scale factor will contract to a minimum value, greater than zero, and then it will expand again. It will 
continue to oscillate, for ever, between maxima and minima values. If we increase the value of $C$, both maxima and minima values will increase. The 
minima values will increase until they reach the value $a_0$. On the other hand, the noncommutative solution Eq. (\ref{29,6}), may describe different 
scale factor behaviors. We studied that solution for several different values of $\gamma$ and $C$ and reached the following conclusions. For positive
values of $\gamma$ and small values of $C$, the NC solutions have the same general behavior than the commutative solution, except that their 
maximum sizes are bigger than the commutative one and they take greater times, than the commutative one, to reach the {\it Big Crunch} singularity. 
%If we increase the absolute values of $\gamma$, that behavior becomes even more pronounced. 
This behavior is qualitatively very similar to the one described by Eq. (\ref{18,6}), 
for $\gamma > 0$, in comparison with the commutative solution Eq. (\ref{19}), for the radiation model Subsection \ref{subsec:radiationk1}. 
An example to this case will produce a figure, qualitatively, very similar to Figure 1. If we 
increase the values of $C$, for different values of $\gamma > 0$, the NC scale factors Eq. (\ref{29,6}) still have maxima values greater than the 
commutative solution but now they reach the {\it Big Crunch} singularity first than the commutative solution. 
%An example of this case is shown in Figure 10. 
Finally, for sufficiently large values of $C$
and different values of $\gamma > 0$, the NC scale factors Eq. (\ref{29,6}) still have maxima values greater than the ones in the commutative case but now they will
contract to minima values, greater than zero and smaller than in the commutative case, and then they will expand again. They will continue to oscillate, 
for ever, between maxima and minima values. If we increase the value of $C$, both maxima and minima values will increase. The minima values will increase 
until they reach the value $a_0$. The maxima and minima values of $a_{nc}$ Eq. (\ref{29,6}) will always be greater and smaller, respectively, than in the 
commutative solution Eq. (\ref{30}). The NC and commutative curves cross at $t = n\pi$, where $n=1,2,..$. An example of this cases is shown in Figure 10.

For negative $\gamma$, we found several differences with respect to the cases where $\gamma >0$. For small $C$, the $a_{nc}$ Eq. (\ref{29,6}) has a
maximum smaller than the commutative one and reach the {\it Big Crunch} singularity before the commutative one. 
This behavior is qualitatively very similar to the one described by Eq. (\ref{18,6}), 
for $\gamma < 0$, in comparison with the commutative solution Eq. (\ref{19}), for the radiation model Subsection \ref{subsec:radiationk1}. 
An example to this case will produce a figure, qualitatively, very similar to Figure 2.
If we increase the values of $C$, the $a_{nc}$ 
still have maxima values smaller than the commutative solution but now they reach the {\it Big Crunch} singularity after the commutative solution.
An example of this case is shown in Figure 11.
For sufficiently large values of $C$, $a_{nc}$ will also oscillate, for ever, between maxima and minima values. The difference with respect to the case
where $\gamma > 0$ is that, here, the maxima and minima values of $a_{nc}$ will always be smaller and greater, respectively, than in the commutative solution. 
Here, also, the NC and commutative curves cross at $t = n\pi$, where $n=1,2,..$.

The results described above will be valid for any values of $a_0$. The different values of $C$, in each domain of that constant, described above, will depend 
on the value of $v_0$. The greater $v_0$ the greater is the value of $C$ in each domain.

\begin{figure}
\includegraphics[width=6cm,height=6cm,angle=0]{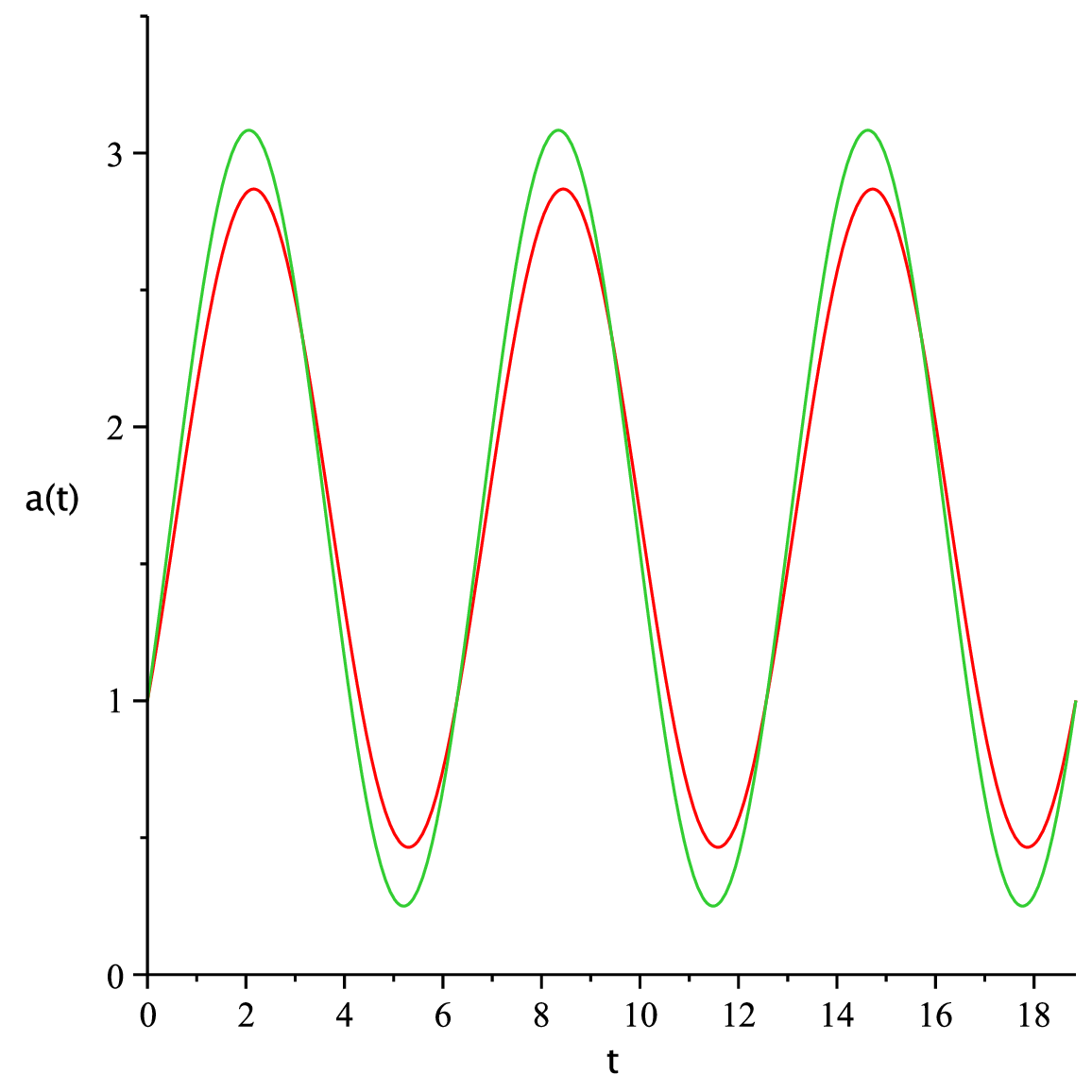}
\caption{{\protect\footnotesize {Scale factor behavior for both commutative and noncommutative cases, when $k=1$. Here, we consider
$\gamma=0.5$, $C=20$ $a_0=1$ and $v_0=1$. The $a(t)$ for the commutative case is the one that has smaller maxima and greater minima.}}}
\label{fig7}
\end{figure}

\begin{figure}
\includegraphics[width=6cm,height=6cm,angle=0]{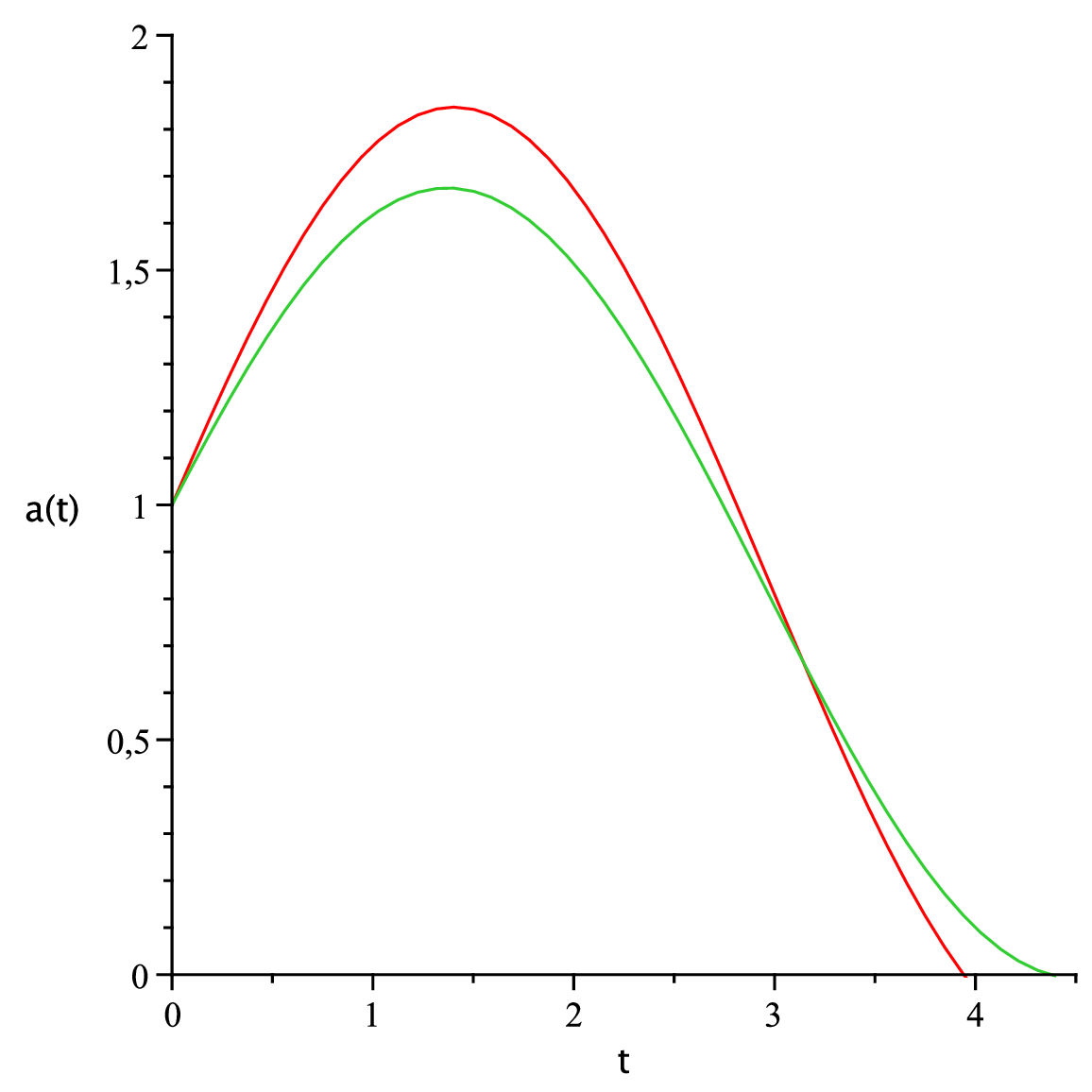}
\caption{{\protect\footnotesize {Scale factor behavior for both commutative and noncommutative cases, when $k=1$. Here, we consider
$\gamma=-0.35$, $C=10$ $a_0=1$ and $v_0=1$. The $a(t)$ for the commutative case is the one that reach $a=0$ first.}}}
\label{fig8}
\end{figure}

\subsection{The case $k = -1$}
\label{subsec:dustk-1}

If we set $k=-1$ in the system Eqs. (\ref{22})-(\ref{23}), we may integrate it to find the following algebraic solutions for $a_c(t)$,
\ba
\label{31}
a_c(t) &=& \left(a_0 + \frac{\gamma}{2}(- v_0 + T_0) + \frac{C}{12}\right)\cosh(t)\\\nonumber
&+& \left(v_0 - \frac{\gamma C}{24}\right)\sinh(t) + \frac{\gamma}{2}\left(\frac{C}{12}t + v_0 - T_0\right) - \frac{C}{12},
\ea
and $T_c(t)$,
\ba
\label{31,5}
T_c(t) &=& \left(a_0 + \frac{\gamma}{2}(v_0 + T_0) + \frac{C}{12}\right)\sinh(t)\nonumber\\
&+& \left(v_0 + \gamma\left(\frac{C}{24} + a_0\right)\right)\cosh(t)\nonumber\\
&+& \gamma\left(\frac{C}{24} - a_0\right) - \frac{C}{12}t - v_0 + T_0.
\ea
Now, we can compute $a_{nc}$ Eq. (\ref{5,55}), with the aid of $a_c(t)$ Eq. (\ref{31}) and $T_c(t)$ Eq. (\ref{31,5}).
It is given, to first order in $\gamma$, by,
\ba
\label{32}
a_{nc}(t) &=& \left(a_0 + \frac{\gamma}{2}T_0 + \frac{C}{12}\right)\cosh(t)\nonumber\\
&+& \left(v_0 + \frac{\gamma}{2}a_0\right)\sinh(t) - \frac{C}{12}.
\ea
If we set $\gamma=0$ in Eqs. (\ref{31}) or (\ref{32}), we obtain the scale factor in the usual commutative model,
\be
\label{33}
a_{c,\gamma=0}(t) = \left(a_0 + \frac{C}{12}\right)\cosh(t) + v_0\sinh(t) - \frac{C}{12}.
\ee
Now, we would like to compare the NC scale factor Eq. (\ref{32}) with the usual commutative one $a_{c,\gamma=0}(t)$ Eq. (\ref{33}).
We shall set $T_0=0$ in $a_{nc}(t)$ Eq. (\ref{32}).

The commutative solution Eq. (\ref{33}) describes a universe that starts from $a_0$ at $t=0$ and expands exponentially to an infinity size 
when the time goes to infinity. Which is the same behavior described by the commutative solution of the radiation model Eq. (\ref{21}),  
Subsection \ref{subsec:radiationk-1}. On the other hand, the noncommutative solution Eq. (\ref{32}), may describe different scale
factor behaviors. We studied that solution for several different values of $\gamma$ and $C$ and reached the following conclusions.
For positive $\gamma$, the noncommutative solutions have the same general behavior than the commutative solution, 
except that they expand in a faster rate than that solution. If we increase the absolute values of $\gamma$, that behavior becomes even more 
pronounced. Therefore, we conclude that when $\gamma$ is positive, the resulting effect upon the scale factor dynamics is the appearance of an 
additional repulsive force, compared to the commutative case. An example to this case will produce a figure, qualitatively, very similar to Figure 3. 
For negative $\gamma$ and $v_0 > |\gamma/2|a_0$, the NC scale factor has the same general behavior than the commutative solution, except that 
it expands in a slower rate than that solution. An example to this case will produce a figure, qualitatively, very similar to Figure 4.
For negative $\gamma$ and $v_0 < |\gamma/2|a_0$, the NC scale factor starts contracting from $a_0$ until it reaches a minimum value. Then, it
expands exponentially to infinity in a slower rate than the commutative solution. An example to this case will produce a figure, qualitatively, very similar to Figure 5.
If we increase the absolute values of $\gamma$ and $a_0$, that behavior becomes even more pronounced. Therefore, we conclude that when $\gamma$ 
is negative, the resulting effect upon the scale factor dynamics is the appearance of an additional attractive force, compared to 
the commutative case. If we increase the value of $C$ both commutative and noncommutative solutions will expand in a faster rate. But all
other properties of those solutions, described above, will remain the same.

\section{Comparison with a model which has a different choice of deformed Poisson brackets}
\label{sec:comparison}

As mentioned, above, in Ref. \cite{gil} the authors considered a very similar classical, noncommutative, FRW model coupled to a perfect fluid, 
in the presence of a cosmological constant. The only differences between the present NC model and the one in Ref. \cite{gil} are the choices of deformed PBs and the 
presence of a cosmological constant in their model. Therefore, in this section, we shall compare both models and verify if they lead to the 
same results or not.

In Ref. \cite{gil}, the authors introduced the noncommutativity, in the model, using a NC generalized symplectic formalism. As we shall see,
this procedure is equivalent to perform transformations, like Eqs. (\ref{5}), leading from the NC variables to the 
commutative ones plus a NC parameter. The starting point of both NC models is considering that the total Hamiltonian has the same functional 
form as Eq. (\ref{3}). Therefore, initially, it will have the same expression as the total Hamiltonian Eq. (\ref{3,5}). In order to be able to 
compare both NC models, we shall simplify the model introduced in Ref. \cite{gil} by setting the cosmological constant to zero. In Ref. \cite{gil}, 
the noncommutativity is introduced by the following choice of deformed Poisson brackets,
\ba
\label{35}
\left\{a_{nc} , T_{nc}\right\} = \theta,\, \left\{P_{anc} , P_{Tnc}\right\} = \beta,\nonumber\\
\left\{a_{nc} , P_{anc}\right\} = \left\{T_{nc} , P_{Tnc}\right\} = 1,\nonumber\\
\left\{a_{nc} , P_{Tnc}\right\} = \left\{T_{nc} , P_{anc}\right\} = 0,
\ea 
which may be compared to our choice given in Eq. (\ref{4}). In fact, although the authors of Ref. \cite{gil} develop the NC generalized
symplectic formalism for both deformed PBs introduced in Eq. (\ref{35}), when they write the resulting total Hamiltonian with the commuting
variables plus the NC parameters, they set $\theta=0$. Therefore, we shall restrict our attention to the NC model described by
the deformed PBs Eq. (\ref{35}), with $\theta=0$. In order to describe this model in terms of usual commutative variables,
which satisfy the usual PBs, we introduce the following transformations from NC variables to commutative ones,
\begin{eqnarray}
\label{36}
P_{anc} & \rightarrow & P_{ac} + \beta T_c,\nonumber\\
a_{nc} & \rightarrow & a_c,\, T_{nc}\rightarrow T_c,\, P_{Tnc}\rightarrow P_{Tc}.
\end{eqnarray}
If one introduces the NC variables Eq. (\ref{36}), in the deformed PBs Eq. (\ref{35}), with $\theta=0$, and uses the usual PBs among the commutative 
variables, it is not difficult to show that they are satisfied to first order in $\beta$. Observing the transformations Eqs. (\ref{36}), we notice the 
first difference between both models. Here, the $a_{nc}$ coincides with $a_c$, therefore the dynamics of the models is described by $a_c$. On the other
hand, we saw that in our model the dynamics is described by $a_{nc}$ Eq. (\ref{5,55}), which does not coincide with $a_c$. 
Now, we rewrite the total Hamiltonian $N_{nc} {\mathcal{H}}_{nc}$ (\ref{3,5}), in terms of the commutative variables Eq. (\ref{36}),
\begin{equation}
N_{nc} {\mathcal{H}}_{nc} = -\frac{(P_{ac} + \beta T_c)^2}{24} - 6ka_c^2 + a_c^{1-3\alpha}P_{Tc}. 
\label{37}
\end{equation}
In order to better compare both NC models, we decided to write the model of Ref. \cite{gil} in the gauge ($N_{nc} = a_{nc}$). It was written
originally in the gauge $N=1$. Apart from this difference and few numerical values, the total Hamiltonian (\ref{37}) coincides with the one obtained in 
Ref. \cite{gil}. Now, we would like to compute the scale factor dynamical behavior, for the present NC model. Initially, we
compute the Hamilton's equations from $N_{nc} {\mathcal{H}}_{nc}$ Eq. (\ref{37}) and, then, combine them to obtain the following second order differential equation,
to first order in $\beta$,
\ba
\label{38}
\ddot a_c(t) & + & ka_c(t) - \frac{C}{12}(1-3\alpha)a_c^{-3\alpha}(t)\nonumber\\
& + & \frac{\beta}{12}(2-3\alpha)a_c^{-3\alpha+1}(t) = 0,
\ea
where $C$ has the same physical meaning as in our NC model. Another difference, between both NC models, is that it is possible to write a Friedmann equation, which depends 
only on $a_c(t)$ and its first time derivative, for the present NC model. We shall not write it, here, because it is not possible to do it for our NC model.
Equation (\ref{38}), is the equivalent, in the present NC model, to the system
of coupled differential equations (\ref{13})-(\ref{14}), in our NC model. If we set $\gamma=0$, in the system (\ref{13})-(\ref{14}), they decouple 
and Eq. (\ref{13}) gives the correct commutative second order differential equation to $a_c(t)$. Which is the same equation one obtains, by setting 
$\beta=0$, in Eq. (\ref{38}). Observing Eq. (\ref{38}), we notice another difference between the two NC models. Here, it is possible to write 
a unique second order differential equation for the scale factor, for a general type of perfect fluid. On the other hand, in our NC model it was 
not possible. There, we have the system of two coupled differential equations (\ref{13})-(\ref{14}). Since we cannot find algebraic solutions, for a generic 
perfect fluid, for either the system Eqs. (\ref{13})-(\ref{14}) or the equation (\ref{38}), we shall restrict our comparisons, between the two NC models, 
for the radiation and dust perfect fluids. All comparisons between both NC models are done using the same values of $\beta$ and $\gamma$ and the other 
corresponding parameters in each model.

\subsection{Radiation perfect fluid}
\label{subsec:comparisonradiation}

Setting $\alpha=1/3$ in Eq. (\ref{38}), we obtain,
\be
\label{39}
\ddot a_c(t) + ka_c(t) + \frac{\beta}{12} = 0.
\ee
This equation must be compared to Eq. (\ref{16}). 
%We observe few important differences between the terms that do not depend on the scale factor or its 
%second time derivative, in both equations. In Eq. (\ref{39}) this
%term is a constant and it does not depend on $k$. Whereas in Eq. (\ref{16}), this term is time dependent and depends on $k$ and
%$T_0$. Those differences, will introduce important distinctions between the solutions to those two equations. 
The solutions to Eq. (\ref{39}), for different values of $k$, are given by,
\ba
\label{40}
a_c(t) & = & -\frac{1}{24}\beta t^2 + v_0 t + a_0,\quad k = 0,\\
\label{41}
a_c(t) & = & v_0\sin(t) + \left(a_0 + \frac{\beta}{12}\right)\cos(t) - \frac{\beta}{12},\nonumber\\
       &   & k = 1,\\
\label{42}
a_c(t) & = & v_0\sinh(t) + \left(a_0 - \frac{\beta}{12}\right)\cosh(t) + \frac{\beta}{12},\nonumber\\
       &   & k = -1.
\ea

The first important difference between the scale factor behavior in the two NC models appears for $k=0$. In our NC model 
the solution for $k=0$, $a_{nc}(t)$, is a first order polynomial in $t$ Eq. (\ref{16,6}), whereas, here, it is a second
order one Eq. (\ref{40}). Another important difference is the presence of a minus sign in front of the $\beta$ term in Eq. (\ref{40}).
It means that, in several cases, that solution will have the opposite behavior than $a_{nc}(t)$ Eq. (\ref{16,6}), when $\gamma$ and $\beta$ have the same sign.

For $k=1$, the solution to Eq. (\ref{39})
%, like Eq. (\ref{16}), describes a driven harmonic oscillator, but in this case,
%under the driving force $-\beta/12$. Its solution, 
is given by $a_c(t)$ Eq. (\ref{41}). It must be compared to $a_{nc}(t)$ Eq. (\ref{18,6}). From Eq. (\ref{41}), it
is clear that the scale factor for this NC model describes a Universe that starts expanding from
$a_0$ at $t=0$, then it reaches a maximum size and, finally, collapses to a {\it Big Crunch} singularity. Qualitatively, $a_c(t)$ Eq. (\ref{41}) has the same general 
behavior than $a_{nc}(t)$ Eq. (\ref{18,6}). On the other hand, quantitatively, for $T_0=0$ and any values of $a_0$ and $v_0$, they have some differences.
For $\gamma>0$ and $\beta>0$ with the same values, the maximum value of $a_{nc}(t)$ Eq. (\ref{18,6}) is bigger than the one of $a_c(t)$ Eq. (\ref{41}). 
Also, $a_{nc}(t)$ Eq. (\ref{18,6}) takes a greater time than $a_c(t)$ Eq. (\ref{41}) to return to $a=0$. 
%If we increase the absolute values of $\gamma$ and $\beta$, those behaviors become even more pronounced. 
%An example of these cases is shown in Figure 1. 
On the other hand, for $\gamma$ and $\beta$ with the same values and negatives, the maximum of $a_c(t)$ Eq. (\ref{41}) is bigger than the one of $a_{nc}(t)$ 
Eq. (\ref{18,6}). Also, $a_c(t)$ Eq. (\ref{41}) takes a greater time than $a_{nc}(t)$ Eq. (\ref{18,6}) to return to $a=0$. 
%If we increase the absolute values of $\gamma$ and $\beta$, those behaviors become even more pronounced. 
%An example of these cases is shown in Figure 2.

For $k=-1$, the solution to Eq. (\ref{39}) is given by $a_c(t)$ Eq. (\ref{42}). It must be compared to $a_{nc}(t)$ Eq. (\ref{20,6}). 
From Eq. (\ref{42}), we observe that there are three different possible evolutions for $a_c(t)$ if $\beta>0$. For $\beta<12a_0$, it describes an universe that starts from $a_0$ at $t=0$ 
and, then, expands exponentially to an infinity size when the time goes to infinity. For $\beta=12a_0$ and $v_0=0$, it describes an universe where $a(t)$ is constant and equal to $\beta/12$.
For $\beta>12(a_0+v_0)$, it describes an universe that starts from $a_0$ at $t=0$, expands to a maximum size and, then, contracts to a {\it Big Crunch}. In contrast, as we saw in Subsection
\ref{subsec:radiationk-1}, for any values of $a_0$ and $v_0$, $a_{nc}(t)$ Eq. (\ref{20,6}) describes an universe that starts from $a_0$ at $t=0$ 
and, then, expands exponentially to an infinity size when the time goes to infinity. In fact, when both solutions expand, $a_{nc}(t)$ Eq. (\ref{20,6}) will always expands faster than $a_c(t)$ 
Eq. (\ref{42}), for any values of $a_0$, $v_0$ and $\gamma$=$\beta$. For $\beta<0$, $a_c(t)$ Eq. (\ref{42}), describes an universe that starts from $a_0$ at $t=0$ 
and, then, expands exponentially to an infinity size when the time goes to infinity, for any values of $a_0$ and $v_0$. In contrast, as we saw in Subsection
\ref{subsec:radiationk-1}, for $\gamma<0$ and $v_0 > |\gamma/2|$,
$a_{nc}(t)$ Eq. (\ref{20,6}) has the same general behavior than $a_c(t)$ Eq. (\ref{42}). Except that it expands in a slower rate than that solution, for $\gamma=\beta$.
For $\gamma<0$ and $v_0 < |\gamma/2|$, $a_{nc}(t)$ Eq. (\ref{20,6}) starts contracting from $a_0$ until it reaches a minimum value. Then, it
expands exponentially to infinity in a slower rate than $a_c(t)$ Eq. (\ref{42}), for $\gamma=\beta$.

\subsection{Dust perfect fluid}
\label{subsec:comparisondust}

Setting $\alpha=0$ in Eq. (\ref{38}), we obtain,
\be
\label{43}
\ddot a_c(t) + \left(k + \frac{\beta}{6}\right)a_c(t) - \frac{C}{12} = 0,
\ee
%This equation must be compared to Eq. (\ref{25}). We observe few important differences between the terms that do not depend on the scale factor or its second time derivative, in both equations. 
%Although, both of them depend on the, respective, noncommutative parameters, and on the constant $C$, 
%In Eq. (\ref{43}) this term is a constant. Whereas in Eq. (\ref{25}), this term is time dependent and depends on $C_1$. Another difference between Eqs. (\ref{25}) and
%(\ref{43}), is that, the term that depends on the scale factor is modified in Eq. (\ref{43}), due to the NC parameter $\beta$.
%Those differences, will introduce important distinctions between the solutions to those two equations. Using the initial
%conditions Eqs. (\ref{17}), the solutions to Eq. (\ref{43}) are given by,
The solutions to Eq. (\ref{43}), for different values of $k$, are given by,
\ba
\label{44}
& a_c(t) & = \sqrt{\frac{6}{\beta}}v_0\sin{\left(\sqrt{\frac{\beta}{6}}t\right)} + \left(a_0 - \frac{C}{2\beta}\right)\cos{\left(\sqrt{\frac{\beta}{6}}t\right)}\nonumber\\
& + & \frac{C}{2\beta},\, k = 0.\\
\label{45}
& a_c(t) & = \frac{v_0}{\sqrt{1+\frac{\beta}{6}}}\sin{\left(\sqrt{1+\frac{\beta}{6}}t\right)} + \frac{C}{12+2\beta}\nonumber\\ 
& + & \left(a_0 - \frac{C}{12+2\beta}\right)\cos{\left(\sqrt{1+\frac{\beta}{6}}t\right)},\, k = 1.\\
\label{46}
& a_c(t) & = \frac{v_0}{\sqrt{1-\frac{\beta}{6}}}\sinh{\left(\sqrt{1-\frac{\beta}{6}}t\right)} - \frac{C}{12-2\beta}\nonumber\\ 
& + & \left(a_0 + \frac{C}{12-2\beta}\right)\cosh{\left(\sqrt{1-\frac{\beta}{6}}t\right)},\, k = -1.
\ea

The first important difference between the scale factor behavior in the two NC models appears for $k=0$. 
For $\gamma>0$ in our NC model, $a_{nc}(t)$ Eq. (\ref{28,6}) describes an universe that originates in $a_0$ at $t=0$ 
and then grows to an infinite size in an infinite period of time. In contrast, for $\beta>0$, $a_c(t)$ Eq. (\ref{44}) describes an universe that also starts from $a_0$ at $t=0$, but then 
expands to a maximum size and finally collapses to a {\it Big Crunch}. For $\gamma$ and $\beta$ negatives, both $a_c(t)$ Eq. (\ref{44}) and $a_{nc}(t)$ Eq. (\ref{28,6}),
describe universes that expand exponentially to infinity. But $a_c(t)$ Eq. (\ref{44}) always expands faster than $a_{nc}(t)$ Eq. (\ref{28,6}).
All behaviors are valid for any values of $a_0$, $v_0$ and $C$.

For $k=1$, the solution to Eq. (\ref{43}) is given by $a_c(t)$ Eq. (\ref{45}). It must be compared to $a_{nc}(t)$ Eq. (\ref{29,6}).
$a_c(t)$ Eq. (\ref{45}) describes an universe that starts expanding from $a_0$ at $t=0$, expands to a maximum size and then
contracts. For small values of $C$ it will contract to a {\it Big Crunch} singularity. For sufficiently large values of $C$ the scale 
factor will contract to a minimum value, greater than zero, and then it will expand again. It will continue to oscillate, for ever, 
between maxima and minima values. Qualitatively, $a_c(t)$ Eq. (\ref{45}) has the same general behavior than $a_{nc}(t)$ Eq. (\ref{29,6}).
On the other hand, quantitatively, for $T_0=0$ and any values of $a_0$ and $v_0$, they have some differences. For small values of $C$,
$\gamma>0$ and $\beta>0$ with the same values, $a_{nc}(t)$ Eq. (\ref{29,6}) has maxima bigger than $a_c(t)$ Eq. (\ref{45}) and it takes greater times than 
$a_c(t)$ Eq. (\ref{45}), to reach the {\it Big Crunch} singularity. Another difference appears for small values of $C$, $\gamma<0$ and $\beta<0$ with the same values.
Now, $a_{nc}(t)$ Eq. (\ref{29,6}) has maxima smaller than $a_c(t)$ Eq. (\ref{45}) and it takes smaller times than $a_c(t)$ Eq. (\ref{45}), 
to reach the {\it Big Crunch} singularity. Another important difference between those NC solutions appears for sufficiently large values of $C$ and
$\gamma$ and $\beta$ of any sign. They oscillate between maxima and minima values with different frequencies. The frequency of $a_c(t)$ 
Eq. (\ref{45}) is modified due to the NC parameter $\beta$.

For $k=-1$, the solution to Eq. (\ref{43}) is given by $a_c(t)$ Eq. (\ref{46}). It must be compared to $a_{nc}(t)$ Eq. (\ref{32}). For any values of $\beta$, $a_0$, $v_0$ and $C$,
$a_c(t)$ Eq. (\ref{46}) describes an universe that starts from $a_0$ at $t=0$ and, then, expands exponentially to an infinity size when the time goes to infinity. For $\gamma>0$,
$a_{nc}(t)$ Eq. (\ref{32}), have the same qualitative behavior as it was shown in Subsection \ref{subsec:dustk-1}. In contrast, quantitatively, it always expands in a faster rate
than $a_c(t)$ Eq. (\ref{46}), for the same values of $\beta>0$ and $\gamma>0$. For negative $\gamma$, any value of $a_0$ and $v_0 > |\gamma/2|a_0$, the $a_{nc}(t)$ Eq. (\ref{32}) 
has the same general behavior than $a_c(t)$ Eq. (\ref{46}), except that it expands in a slower rate than that solution, for the same values of $\gamma<0$ and $\beta<0$. For negative 
$\gamma$, any value of $a_0$ and $v_0 < |\gamma/2|a_0$, the $a_{nc}(t)$ Eq. (\ref{32}) starts contracting from $a_0$ until it reaches a minimum value. Then, it expands exponentially 
to infinity in a slower rate than $a_c(t)$ Eq. (\ref{46}), for the same values of $\gamma<0$ and $\beta<0$.

\section{Conclusions}
\label{sec:conclusions}

We conclude that the noncommutativity greatly modifies the original commutative cosmological model. Since we are particularly interested in describing the
present expansion of our Universe, we may mention that, due uniquely to the noncommutativity introduced here, we obtained scale factor solutions compatible with
that expansion. In the dust NC model, for $k=0$, we obtained a scale factor solution describing
an exponential expansion, not present in the corresponding commutative solution. Since, the dust perfect fluid represents the present matter dominated era of
our Universe and $k=0$ is a good candidate to describe its spatial curvature, then, under those circumstances, our NC model may be considered a possible candidate
to describe the present expansion of the Universe. On the other hand, if the observations establish that the Universe is better described by a model with negative 
spatial curvature, we may mention that, in both radiation and dust NC models, for $k=-1$, we obtained scale factor solutions describing exponential expansions. 
Although, in this case, the corresponding commutative solutions describe also exponential expansions, in the NC models we have free parameters, not present in the 
commutative models, that may better adjust the observational data. 
%Besides the solutions described above, compatible with the present expansion of the Universe,
%we also obtained expansive solutions in both radiation and dust NC models, for $k=1$. For those solutions the scale factor expands linearly in time, oscillating
%between maximum and minimum values. That type of expansive solution is not present in the corresponding commutative models.

Another important conclusion comes from our comparison with another NC cosmological model \cite{gil}. The only difference between the two models is how the noncommutativity
is introduced. From this comparison we conclude that because the noncommutativity is introduced there, through different deformed Poisson brackets between the variables, 
than here, the cosmological models give different dynamical scale factor equations and predictions. 
Consider as an example, the dust case with $k=0$. If $\gamma>0$, our NC model describes an universe that originates in $a_0$ at $t=0$ and then grows to an infinite size in 
an infinite period of time. In contrast, if $\beta>0$, the NC model of Ref. \cite{gil} describes an universe that also starts from $a_0$ at $t=0$, but then 
expands to a maximum size and finally collapses to a {\it Big Crunch}.
We believe that only through observations one may verify whether
NC is important, or not, in order to describe our Universe. If one concludes that NC is indeed important, those observations would, also, establish which way of introducing the 
noncommutativity is more appropriate.

\begin{acknowledgements}
%G. Oliveira-Neto thanks FAPEMIG for partial financial support. 
L. G. Rezende Rodrigues e M. Silva de Oliveira thank CAPES for their scholarships.
G. A. Monerat thank UERJ for the Proci\^{e}ncia grant, via FAPERJ.
\end{acknowledgements}

\end{document}